\documentclass[sigconf]{acmart}

\usepackage{ifthen}

\definecolor{gray}{rgb}{0.8,0.8,0.8} 
\definecolor{green}{rgb}{0, 0.6, 0} 
\definecolor{orange}{rgb}{1, 0.5, 0} 	
\definecolor{mahogany}{rgb}{0.75, 0.25, 0.0}
\definecolor{purple}{rgb}{0.6, 0, 0.6}
\definecolor{darkgreen}{rgb}{0, 0.4, 0}
\definecolor{lightblue}{rgb}{0, 0.8, 1} 
\definecolor{lightgreen}{rgb}{0.9, 1, 0.9} 
\definecolor{lightred}{rgb}{1, 0.3, 0.3} 
\definecolor{navy}{rgb}{0, 0, 0.4}
\definecolor{highlight}{rgb}{1.0, 0.13, 0.32}
\definecolor{black}{rgb}{0.0, 0.0, 0.0}

\provideboolean{revising}
\setboolean{revising}{false}

\ifthenelse{\boolean{revising}}
{
    \newcommand{\revision}[1]{\textcolor{blue}{#1}}
    \newcommand{\delete}[1]{\textcolor{gray}{\sout{#1}}}
    \newcommand{\deletesection}[1]{\textcolor{gray}{{#1}}}

}{
	\newcommand{\revision}[1]{#1}
	\newcommand{\delete}[1]{}
        \newcommand{\deletesection}[1]{}

    \newcommand{\tc}[1]{}
        
}

\usepackage{multirow}
\usepackage{subcaption} 
\usepackage[table]{xcolor}
\usepackage{enumitem}
\newlist{tightitemize}{itemize}{1}
\setlist[tightitemize]{nosep,left=0pt,label=\textbullet}

\usepackage{siunitx}
\AtBeginDocument{%
  \providecommand\BibTeX{{%
    \normalfont B\kern-0.5em{\scshape i\kern-0.25em b}\kern-0.8em\TeX}}}

\copyrightyear{2026}
\acmYear{2026}
\setcopyright{cc}
\setcctype{by}
\acmConference[CHI '26]{Proceedings of the 2026 CHI Conference on Human Factors in Computing Systems}{April 13--17, 2026}{Barcelona, Spain}
\acmBooktitle{Proceedings of the 2026 CHI Conference on Human Factors in Computing Systems (CHI '26), April 13--17, 2026, Barcelona, Spain}
\acmPrice{}
\acmDOI{10.1145/3772318.3790897}
\acmISBN{979-8-4007-2278-3/2026/04}

\begin{document}


\title{Designing Staged Evaluation Workflows for LLMs: Integrating Domain Experts, Lay Users, and Model-Generated Evaluation Criteria}




\author{Annalisa Szymanski}
\email{aszyman2@nd.edu}
\affiliation{
\department{Department of Computer Science and Engineering}
\institution{University of Notre Dame}
\city{Notre Dame}
\state{IN}
\country{USA}}

\author{Simret Araya Gebreegziabher}
\email{sgebreeg@nd.edu}
\affiliation{
\department{Department of Computer Science and Engineering}
\institution{University of Notre Dame}
\city{Notre Dame}
\state{IN}
\country{USA}}

\author{Oghenemaro Anuyah}
 \email{maroanuyah@microsoft.com}
\affiliation{%
 \institution{Microsoft Corporation}
 \city{Redmond}
 \state{WA}
 \country{USA}
 }
 
\author{Ronald A. Metoyer}
\email{rmetoyer@nd.edu}
\affiliation{
\department{Department of Computer Science and Engineering}
\institution{University of Notre Dame}
\city{Notre Dame}
\state{IN}
\country{USA}}

\author{Toby Jia-Jun Li}
\email{toby.j.li@nd.edu}
\affiliation{
\department{Department of Computer Science and Engineering}
\institution{University of Notre Dame}
\city{Notre Dame}
\state{IN}
\country{USA}}

%
\renewcommand{\shortauthors}{Szymanski et al.}


\begin{abstract}
Large Language Models (LLMs) are increasingly utilized for domain-specific tasks, yet evaluating their outputs remains challenging. A common strategy is to apply \textit{evaluation criteria} to assess alignment with domain-specific standards, yet little is understood about how criteria differ across sources or where each type is most useful in the evaluation process. This study investigates criteria developed by domain experts, lay users, and LLMs to identify their complementary roles within an evaluation workflow.  Results show that experts produce fact-based criteria with long-term value, lay users emphasize usability with a shorter-term focus, and LLMs target procedural checks for immediate task requirements. We also examine how criteria evolve between \textit{a priori} and \textit{a posteriori} phases, noting drift across stages as well as convergence in the \textit{a posteriori} phase. Based on our observations, we propose \revision{design guidelines for} a staged evaluation workflow combining the complementary strengths of these sources to balance quality, cost, and scalability.

\end{abstract}

%
%



%
%
\maketitle
\section{Introduction}

Large Language Models (LLMs) have been increasingly utilized across various domains, such as healthcare, nutrition, education, and other specialized fields, to perform complex tasks that require expert-level and evidence-based reasoning~\cite{sarkar2024llms, sallam2023chatgpt, ahmad2023creating, chatelan2023chatgpt, ayers2023comparing}.
Despite their widespread use, significant concerns remain about the accuracy, reliability, and variability in LLM performance, where errors and hallucinations can have serious implications through the perpetuation of biases and potentially harmful misinformation~\cite{andriopoulos2023augmenting, liu2024exploring, kalai2024calibrated, wang2023assessing}. Addressing these risks requires robust evaluation processes that provide developers with insights on how to select appropriate models, assess task suitability~\cite{liusie2024efficient}, and ensure that outputs align with end-user needs~\cite{liu2023trustworthy, huang2024trustllm, gebreegziabher2024supporting}, while also promoting transparency and accountability in model behavior~\cite{liao2023ai}.

However, ensuring that LLM evaluations are effective for complex tasks requires accounting for the unique requirements of each specific domain to accurately assess output quality. Traditional metrics such as BLEU~\cite{papineni2002bleu} and ROUGE~\cite{lin2004rouge} often fail to capture the subjective and context-dependent nature of complex tasks~\cite{pan2024human}. Due to this, human evaluation has become the gold standard~\cite{sun2024lalaeval}, but it also presents many challenges, including cost, time, and scalability~\cite{gehrmann2023repairing}.

The use of \textit{evaluation criteria}, or user-defined assertions to assess the quality of model outputs, has become a widely adopted foundation for model evaluation and now serves as the basis for methods such as LLM-as-a-Judge and LLM juries, which are prevalent in both academic research and industry practice~\cite{Wiltberger_2025, ramnath2025amulet, shankar2024validates, arawjo2024chainforge, kahng2024llm, wang2024prompt}. Evaluation criteria are used as declaration statements to assess specific aspects of the output (e.g., accuracy, relevance, and personalization, etc.) to then be evaluated by LLMs using true/false judgments or measured with ordinal metrics. Tools such as EvalAssist~\cite{desmond2025evalassist}, EvalLM~\cite{kim2024evallm}, or EvalGen~\cite{shankar2024validates} have been used to assist users in generating and refining evaluation criteria based on their assessments, with the aim to better align LLMs to human preferences~\cite{arawjo2024chainforge, shankar2024validates, kim2024evallm, gebreegziabher2024leveraging}. Such criteria are particularly valuable for evaluating outputs in complex tasks, as they enable the integration of diverse domain knowledge and perspectives.

While current approaches support creating metrics and criteria that align LLMs with end-user needs, a critical gap remains in understanding the types of evaluation criteria produced by different sources, as well as when and where each can be useful in the evaluation process. For example, domain experts have the ability to address ethical concerns and contribute domain-specific knowledge, but are often excluded from evaluation stages due to the time and cost required for reviewing outputs~\cite{jiao2024navigating, gururangan2020don}. Lay users, or typical end-users of a domain service, have shown that the human-in-the-loop evaluation process is valuable to reflect real-world needs; however, they are less technically specialized~\cite{wang2024understanding}. LLMs themselves have also been used to generate evaluation criteria and could offset the scalability challenges of integrating humans into the evaluation process~\cite{dong2024can, zheng2024judging}. Comparing the contributions and limitations of these three sources is essential for determining when and how each should be involved to design evaluation workflows that are both high-quality and cost-efficient.

In our study, we aim to examine the alignment between the types of criteria created by domain experts, lay users, and LLMs to better determine the potential roles each may have in the evaluation workflow. Building on previous research that introduced the concept of \textit{``criteria drift''}~\cite{shankar2024validates}, or the evolution of evaluation criteria as users transition from creating them based on prompts \textit{(a priori)} to refining them after reviewing outputs \textit{(a posteriori)}~\cite{shankar2024validates}, we also design our study to investigate how these criteria may change across different phases. Our research is guided by the following questions:

\begin{itemize}
    \item \textbf{(RQ1)} What are the similarities and differences in evaluation criteria among the three sources of domain experts, lay users, and LLMs themselves?  
    \item \textbf{(RQ2)} How do \textit{a priori} (prompt-based) and \textit{a posteriori} (output-based) phases influence the development of evaluation criteria within each source?
    \item \textbf{(RQ3)} What are the design implications of involving different sources to improve the LLM evaluation workflow?
\end{itemize}

To answer our research questions, we conduct a case study in the two specialized fields of nutrition and math pedagogy, which are similar in their need for domain-specific expertise and real-world decision-making, but differ in the types of knowledge applied and the forms of reasoning required to assess LLM outputs. Experts and lay users from each domain establish evaluation criteria for a set of tasks, which we then compare to criteria generated by LLMs.
To analyze how these criteria may evolve, all sources create evaluation criteria both \textit{a priori} and \textit{a posteriori} to allow for the examination of shifts in criterion development and determine the best place to incorporate expertise.

Based on our qualitative analysis, we identified key differences in how evaluation criteria were developed by domain experts, lay users, and LLMs, which suggest where each source may be most strategically integrated within an evaluation workflow.
Domain experts create fact-based, instructional criteria aimed at preventing misconceptions and promoting long-term educational value, often generating these upon reading the prompt, which points to their strength in the \textit{a priori} phase of evaluation. In contrast, lay users emphasize usability and presentation, while LLMs focus on low-level procedural checks tied to immediate task requirements.
We also found that domain experts and lay users create criteria in response to errors and gaps in the output, respectively, while LLMs do not. 
In addition, our results confirm the existence of criteria drift~\cite{shankar2024validates}, and also reveal convergence in the \textit{a posteriori} phase, as all sources created criteria based on features explicitly present in the output.
This raises the question of where human input is most valuable in the evaluation workflow and demonstrates the risk of relying solely on LLMs, which may reinforce errors in outputs and introduce bias or overfitting to flawed responses\revision{, suggesting that improvements in LLM development and prompting strategies are necessary.}
By examining how criteria evolve between the \textit{a priori} and \textit{a posteriori} phases, we can inform more efficient evaluation workflows and guide decisions about when and how to involve each source. These findings motivate a staged evaluation workflow that strategically combines the complementary strengths of domain experts, lay users, and LLMs to balance quality, cost, and scalability.



Our research makes the following contributions to the development of evaluation systems in the HCI field: 

\begin{itemize}
    \item  We provide an empirical understanding of the differences in evaluation metrics developed by domain experts, lay users, and LLMs in \textit{a priori} and \textit{a posteriori} phases.
    \item We identify implications for involving each source, including where human input adds the most value, the risks of relying solely on LLMs \revision{and the influence of prompting strategies on LLM-based criteria generation}, and how criteria evolve through drift and convergence.
    \item \delete{We propose a staged evaluation approach}\revision{We present design guidelines for a staged evaluation approach} that uses the complementary strengths of domain experts, lay users, and LLMs.
\end{itemize}

\section{Background and Related Work}

We begin this section by discussing the importance of evaluation criteria in complex domains to ensure trust and reliability in LLM outputs. We then review existing research on systems developed for LLM evaluations. Finally, we explore the gaps in using diverse perspectives and expertise to improve evaluation processes and refine LLMs for real-world applications.

\subsection{Evaluating LLMs for Complex Tasks}
Although LLMs show significant promise in complex fields due to their vast training data, they are susceptible to biased content, hallucinations, and misinformation, which can limit their practical usefulness and potentially cause harm to end-users~\cite{ji2023survey, huang2023survey}. This is due to the nature of the open-ended prompt questions that require knowledge and logical reasoning capabilities~\cite{lu2024ahp}.
Unlike close-ended questions, where responses are constrained to predefined choices or structured answers (multiple-choice or true/false questions), open-ended prompts require LLMs to generate creative, contextually appropriate, and often multi-faceted outputs~\cite{han2023robustqa, zhu2021retrieving}. 
Due to their statistical nature, LLMs can produce wide variability in content across outputs, raising concerns about the consistency and reliability of the information they generate.   
These complexities introduce greater variability in LLMs, as valid responses are vast and depend on the specific context, task, and user intent~\cite{lu2024ahp}. 

Because the evaluation of these tasks is difficult~\cite{khashabi2021genie, gao2024llm}, recent research in Natural Language Processing (NLP) has focused on empowering LLMs to better understand and follow instructions that align with human preferences~\cite{gabriel2022challenge}. This alignment allows the models to behave more in accordance with human values, intentions, and goals, and has been shown to be critical before deploying LLMs in real-world applications~\cite{liu2023trustworthy, kirk2024benefits}. Studies have shown that models tailored to match the specific needs and motivations of users not only enhance trust, but also improve adoption, making them safer and more effective tools~\cite{liu2023trustworthy, kirk2024benefits}. \revision{Complementing this, research on human–AI collaboration shows that alignment depends not only on model tuning but on workflows that help humans understand when to rely on AI and when to intervene~\cite{spitzer2025human, schafer2025ai}.}

Evaluation criteria, particularly those developed through assertion statements such as true/false metrics or ordinal ranking scales~\cite{shankar2024validates, zhang2024llmeval}, provide a systematic way to measure how well LLMs align with user preferences and task objectives. 
These criteria can be defined in natural language by users and evaluated by applying LLMs themselves as evaluators, often referred to as ``LLM-as-a-judge'' as a way to offset the costs and scalability challenges of human involvement in evaluation workflows~\cite{shankar2024validates}. 
This approach uses the model’s ability to interpret instructions and assess outputs for alignment with predefined criteria~\cite{arawjo2023chainforge, kim2024evallm}. For example, in a healthcare context, a stakeholder, such as a domain expert, could include evaluation criteria such as \textit{``The generated response aligns with clinical guidelines.''} or \textit{``The patient information is accurately reflected in the recommendation.''}.

In real-world contexts, evaluation criteria have offered several practical benefits. For developers, these criteria provide insights into the strengths and weaknesses of the model and help identify performance gaps to guide decisions on when to refine prompts~\cite{arawjo2023chainforge}, retrain models~\cite{shankar2024validates}, or where expert feedback may be required~\cite{szymanski2024integrating}.
As research continues to explore domain-specific tasks, well-defined evaluation workflows that utilize these criteria are essential to address the challenges posed by biases, misinformation, and the inherent limitations of the statistical models of LLMs~\cite{lu2024ahp}.
For end-users, criteria-driven evaluations build confidence in the system by demonstrating that outputs are validated against clear standards~\cite{liu2023trustworthy, huang2024trustllm}. Automated alignment evaluations, where outputs are continually assessed against pre-defined criteria~\cite{chen2024self}, could empower users to decide when to trust and use outputs and should be used before deploying LLMs for complex tasks. In this study, we explore the use of evaluation criteria to enhance evaluation processes with the aim of establishing a reliable framework for assessing LLM outputs in complex tasks.

\subsection{Systems Currently Used for LLM Evaluations}
To support both programmers and end users in evaluating model performance and aligning LLMs with their needs, several interactive systems have been proposed that facilitate the generation and testing of evaluation criteria. These systems are primarily designed to identify bias in generated responses~\cite{chen2024designing}, iterate and optimize prompts~\cite{kim2024evallm}, and select models~\cite{kahng2024llm, chiang2024chatbot} that best align with user objectives. While their individual goals vary, they share the broader aim of improving alignment between user preferences and LLM outputs~\cite{shankar2024validates}. Most of these systems rely on the LLM to mediate evaluation to ensure that task-specific needs are addressed.

Google Labs has recently introduced Stax, an AI evaluation toolkit that allows users to benchmark and compare models against criteria to understand how an LLM is performing~\cite{Wiltberger_2025}. EvalAssist also provides an interface that allows users to create and test evaluation criteria using an LLM-as-a-judge~\cite{desmond2025evalassist}. Tools such as ChainForge~\cite{arawjo2024chainforge} and Promptfoo~\cite{webster2023promptfoo} similarly enable users to define evaluation criteria for assessing response quality across multiple models. EvaluLLM~\cite{desmond2024evalullm} expands this approach by allowing users to generate evaluation outputs from prompts, select multiple models, and define criteria with custom metrics in natural language. Evallm~\cite{kim2024evallm} supports criteria-driven evaluation to refine prompts, while EvalGen~\cite{shankar2024validates} algorithmically generates evaluation assertions from prompt revision histories.


Together, these systems demonstrate the growing role of validation frameworks in ensuring models are better equipped to meet real-world needs. With system support, the developer can analyze certain aspects of the output and make changes to the prompt or fine-tune the model when necessary. However, despite their utility, these systems have mostly been used by developers, who focus on the technical aspects of LLM outputs and may not be able to determine what is relevant or appropriate within the output's context. 
As Pan et al. highlight, while systems like EvaluLLM enable users to leverage LLMs as customizable judges, there is also a necessity of incorporating diverse human input to ensure alignment with user needs and domain-specific standards~\cite{pan2024human}. 

\revision{In addition, recent work on human–AI workflows similarly demonstrates that complex, multi-step tasks require systems that support shared sensemaking and structured collaboration between humans and AI~\cite{ma2025towards, wang2025aideation}. Despite these advances, how such collaborative principles should be translated into evaluation workflows remains underspecified, prompting calls for clearer guidance for structuring human–AI interaction during evaluation ~\cite{sadek2024guidelines, amershi2019guidelines, van2021towards}. Foundational guidelines emphasize that evaluation processes must support user understanding, set expectations, and calibrate appropriate reliance on AI outputs~\cite{amershi2019guidelines}. Additional work positions human–AI collaboration as an ongoing process of co-management in which humans and LLMs negotiate responsibilities and resolve ambiguity together~\cite{wang2020human, song2024human}.}
\revision{Reinforcing the need for clearer structuring of evaluation workflows, recent empirical studies show that evaluation and decision-making with AI unfold through negotiated processes in which humans delegate, verify, and reinterpret AI contributions depending on contextual cues~\cite{spitzer2025human, schafer2025ai, umbelino2025an}. Furthermore, research on trust and reliance shows that workflows must help users calibrate when to depend on AI and when to intervene~\cite{ashktorab2024trust}.}
\revision{Building on this gap, our study investigates how evaluation criteria differ when developed by domain experts, lay users, and LLMs. This comparative perspective allows us to analyze how stakeholders bring distinct forms of contextual reasoning, interpretive judgment, and efficiency to evaluation tasks to mirror the human–AI collaboration challenges highlighted in prior work. By identifying where these perspectives align or diverge, we aim to inform the design of more reliable evaluation workflows and clarify how multi-stakeholder input should be incorporated when assessing LLMs for complex, real-world tasks.}
\delete{Building on this gap, our study explores how evaluation criteria differ when developed by domain experts, lay users, and LLMs. By examining these diverse perspectives, we investigate how to make LLM evaluations more reliable for complex tasks and how multi-stakeholder input can be integrated into evaluation workflows.}

\subsection{Including Different Perspectives in the Evaluation Process}
Our study aims to examine the skills that different sources can bring to evaluations that may complement developers to derive more reliable outputs. To improve the accuracy of LLMs in specific domains, empirical studies have demonstrated the critical role of domain experts involved in evaluating LLMs.
Prior work has conducted detailed validation processes by domain experts to lead to the creation of design guidelines for prompting LLMs in specialized fields~\cite{szymanski2024integrating}. Expert-evaluated models in specialized domains such as radiation oncology, radiology, and nutrition have shown improved performance by aligning outputs with domain-specific standards and reducing inaccuracies~\cite{degachi2025towards, cheng2023now, king2023introduction, liu2023chatcounselor}. Evaluations by experts also ensure that outputs reflect the context-sensitive knowledge required for high-stakes decision-making~\cite{andriopoulos2023augmenting}. These evaluations not only enhance the reliability and usability of LLMs but also build trust among end-users, particularly in domains where errors can lead to significant consequences~\cite{sarkar2024llms, sallam2023chatgpt}. 
While incorporating domain experts in the evaluation process is valuable, adding the unique needs of the lay user should also be considered. Lay users provide the perspectives as end-users using the LLMs for complex tasks. Research in pedagogy, for instance, has shown that expert-driven strategies can sometimes miss the mark due to ``blind spots'' in user performance or needs~\cite{wang2021seeing}. Including lay users in the evaluation process may not only increase usability and user experience but may also allow developers to create systems that more authentically reflect human needs and expectations~\cite{wu2024aligning}.

However, relying solely on human expertise comes with limitations. While domain experts and lay users could provide invaluable feedback to improve model accuracy, reduce biases, and create specialized tools~\cite{szymanski2024integrating, ziegler2019fine}, the cost and time intensiveness of their involvement~\cite{gehrmann2023repairing}, \revision{as well as the difficulty in finding suitable human subjects~\cite{ahmed2025can}} often limit their integration. 
\revision{Prior work has shown that LLMs can operate orders of magnitude faster and at substantially lower cost than human evaluation. For example, Hymel et al. report performance at roughly 720× the speed of human experts and at a fraction of the cost~\cite{hymel2025analysis}.} Previous studies have demonstrated that LLMs can be used to generate evaluation criteria, typically after users evaluate outputs and provide feedback~\cite{desmond2025evalassist, arawjo2024chainforge, shankar2024validates}. 
However, while this can streamline the evaluation process, relying solely on LLMs for this purpose introduces critical risks. Previous studies have shown that LLMs are often unreliable for evaluating complex tasks, as their judgments may not align with domain expert knowledge~\cite{szymanski2024limitations}. Moreover, LLMs evaluating their own or other model outputs risk perpetuating systemic biases and inaccuracies inherent in their training data~\cite{arawjo2024chainforge}. This suggests the potential of achieving a balance between domain experts, lay users, and LLMs in evaluation workflows.
By identifying where the criteria set by each stakeholder group align or diverge, we aim to inform when and how different perspectives should be incorporated into the evaluation workflow.



\section{Methods}

\begin{figure*}[!t]
\centering
\includegraphics[width=0.88\textwidth]{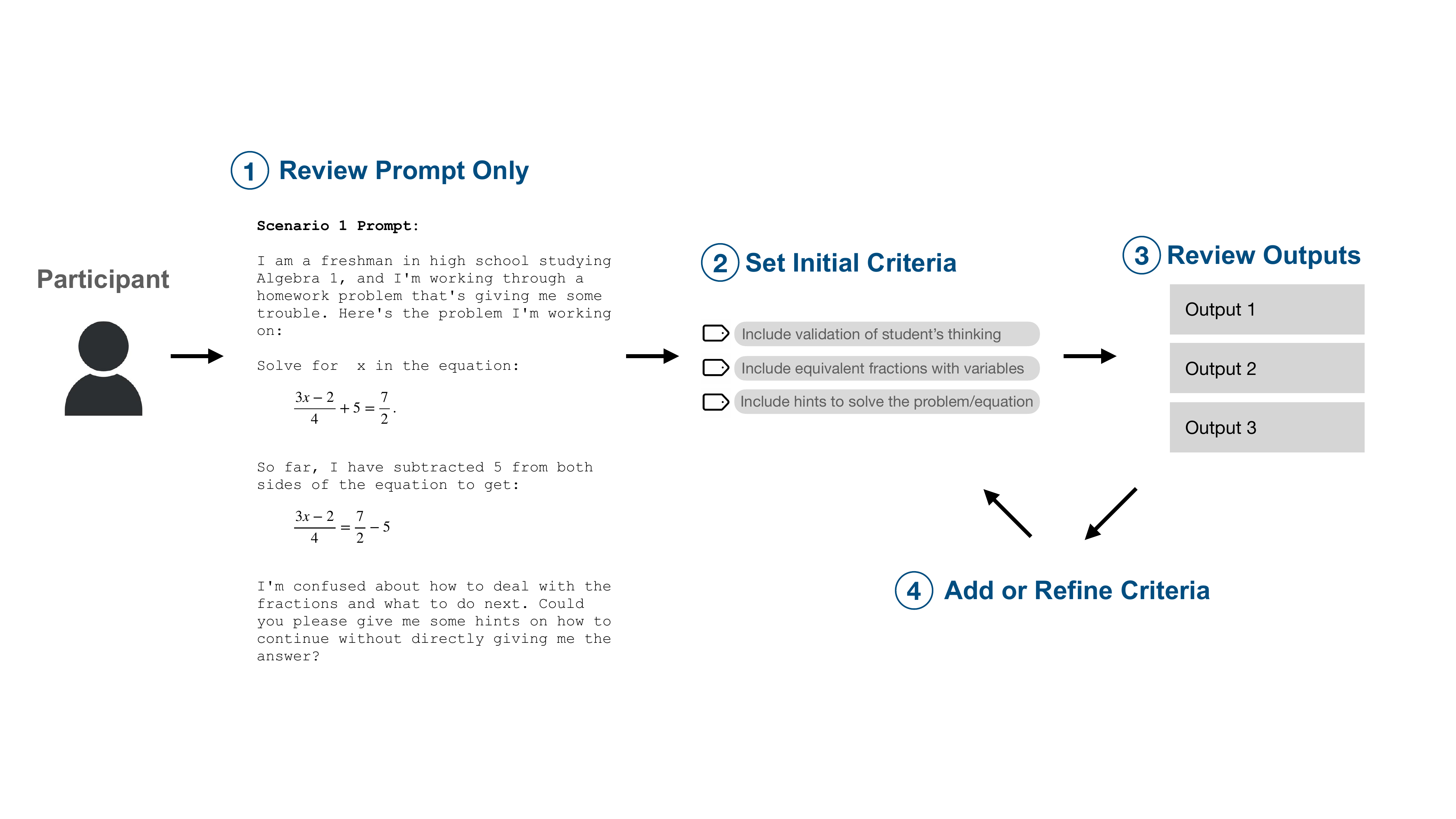}
\caption{The process followed for creating and refining evaluation criteria for LLMs. In step (1), participants (domain experts or lay users) are presented with a prompt to review. In step (2), participants create initial criteria. In step (3), participants review three outputs generated by LLMs. In step (4), each participant subsequently adds or refines criteria based on review of the outputs.}
\Description{The diagram illustrates the four-step process used in our study for evaluating Large Language Model (LLM) outputs. The participant starts by reviewing only the prompt (step 1). The prompt example shown describes a high school student working on an Algebra equation, struggling with fractions. After reviewing the prompt, in step 2, the participant sets initial criteria based on the problem, such as validating the student's thinking, including equivalent fractions with variables, or providing an answer to the problem. In step 3, the participant reviews three outputs generated by the LLM. Finally, in step 4, the participant can add or refine the criteria based on their review of the outputs. }  
\label{fig:process1}
\end{figure*}

We conducted interviews with both domain experts and lay users. Each domain expert developed and refined evaluation criteria for each of three specific scenarios related to their expertise and domains or, in the case of lay users, within the domain where they had sought expert assistance. Simultaneously, we generated criteria using five different LLMs for each of the same three domain-specific tasks for direct comparison. Because of the ``criteria drift'' phenomenon, where human evaluation criteria evolve during the process of assessing LLM outputs~\cite{shankar2024validates}, we structured the criteria creation process in two phases. We were interested to see how the criteria may evolve between first reviewing the prompt only \textit{(a priori)} and then after evaluation of the output \textit{(a posteriori)}. Figure \ref{fig:process1} illustrates the process of criteria development in each of the two phases. The following sections provide detailed information on our methods.

\subsection{Domain Selection}
We selected the domains of nutrition and math pedagogy for their reliance on specialized knowledge and training.
The deployment of LLMs in these fields is increasing, presenting opportunities for significant societal benefits by improving access to critical services~\cite{sonkar2024pedagogical, szymanski2024integrating}. 
In nutrition, registered dietitians undergo specialized education and training to provide evidence-based guidance, particularly for patients with lower health literacy or those managing specific medical conditions~\cite{wood2015exploring}. This domain requires precise evaluation benchmarks from experts to ensure outputs align with professional standards and avoid promoting harmful dietary advice, especially in disease management scenarios~\cite{roseman2021academy}. Similarly, math pedagogy involves expertise in teaching and communicating complex mathematical concepts effectively, with requirements that vary based on the learner’s age, educational context, and alignment with institutional or state standards~\cite{miller2024llm}. Both domains exemplify the complexity of knowledge-driven fields where accurate and reliable LLM outputs can enhance accessibility and support end users in meaningful ways. \revision{While nutrition and math pedagogy both involve structured problem solving, they differ in ways that strengthen the scope of our study. The two domains share similar epistemic structures yet diverge in purpose, stakes, and audience, allowing us to identify patterns in criteria development that are not tied to a single field.}
\newline

\subsection{Participant Recruitment}
Our study received approval from our Institution's Review Board (IRB) prior to participant recruitment. Below, we outline our recruitment efforts for both domain experts and lay users. Both groups were asked to sign a consent form and fill out a pre-interview survey prior to the study. Table \ref{tab:domainexpert} and Table \ref{tab:layuser} outline the demographic and professional profiles of domain experts and lay users, respectively.

\subsubsection{Domain Experts}
For the nutrition domain, we recruited five registered dietitians (RDs) with specialized training and experience in clinical practice. For the pedagogy domain, we recruited five licensed mathematics educators who have taught middle and high school mathematics and hold certifications in specific teaching competencies.

Participants were recruited through email communication with local nutrition practices, schools, and university networks. We then distributed consent forms electronically to all of our participants, which they signed before the study. The registered dietitians had an average of twelve years of professional experience, and mathematics educators averaged eight years of experience. While 60\% of all domain expert participants had prior experience with LLM tools such as ChatGPT or Gemini, none had previously set evaluation criteria for LLMs before participating in our study.

\subsubsection{Lay Users}
To evaluate the differences in criteria development between domain experts and lay users, we recruited participants who had previously received \revision{services in the relevant domain.} \revision{For the nutrition domain, lay users were individuals who had previously received professional guidance from a} dietitian or \delete{nutrition services}\revision{nutritionist for personal health or dietary needs. For the math pedagogy domain, lay users were individuals who, as students, had sought assistance from}\delete{or had consulted with} a tutor or teaching assistant \revision{for learning support}\delete{or other educational support for pedagogical purposes}. The recruitment was carried out through social networks, university networks, and outreach within the local community. 40\% of all lay users participants had used LLM tools, such as ChatGPT or Gemini, for nutrition advice or tutoring purposes, with only one having prior experience in creating evaluation criteria.

\begin{table*}[!t]
\centering
\begin{tabular}{|c|c|c|c|c|c|c|}
\hline
\multicolumn{7}{|c|}{\textbf{Domain Expert Profile}} \\ \hline
\textbf{Domain} & \textbf{\begin{tabular}[c]{@{}c@{}}Domain \\ Expert ID\end{tabular}} & \textbf{\begin{tabular}[c]{@{}c@{}}Years \\ Experience\end{tabular}} & \textbf{\begin{tabular}[c]{@{}c@{}}LLM \\ Familiarity\end{tabular}} & \textbf{Degree} & \textbf{Gender} & \textbf{Specialization} \\ \hline
\multirow{5}{*}{Nutrition} & NutExp1 & 3 & No & Master's Degree & F & Consulting—Private Practice \\ \cline{2-7} 
                           & NutExp2 & 5 & No & Bachelor's Degree & F & Hospital/Medical Practice \\ \cline{2-7} 
                           & NutExp3 & 17 & Yes & Master's Degree & F & Research \& Education \\ \cline{2-7} 
                           & NutExp4 & 11 & No & Bachelor's Degree & F & Community \& Public Health \\ \cline{2-7} 
                           & NutExp5 & 24 & Yes & Master's Degree & F & Hospital/Medical Practice \\ \hline
\multirow{5}{*}{Pedagogy}  & PedExp1 & 7 & No & Bachelor's Degree & F & High School Math \\ \cline{2-7} 
                           & PedExp2 & 6 & Yes & Master's Degree & M & High School Math \\ \cline{2-7} 
                           & PedExp3 & 9 & Yes & Master's Degree & M & High School Math \\ \cline{2-7} 
                           & PedExp4 & 11 & Yes & Bachelor's Degree & F & Middle School Math \\ \cline{2-7} 
                           & PedExp5 & 8 & Yes & Bachelor's Degree & F & Middle School Math \\ \hline
\end{tabular}
\caption{Domain Expert Participant Profile}
\Description{%
A table summarizing background and professional characteristics of domain expert participants across two domains: Nutrition and Math Pedagogy. Each row represents a unique domain expert. Columns include domain, expert identifier, years of professional experience, self-reported familiarity with large language models, highest degree attained, gender, and area of specialization. Nutrition experts span clinical, community, research, and public health contexts, while pedagogy experts specialize in middle and high school mathematics with varied levels of teaching experience.
}
\label{tab:domainexpert}
\end{table*}
\begin{table*}[!t]
\centering
\begin{tabular}{|c|c|c|c|c|c|}
\hline
\multicolumn{6}{|c|}{\textbf{Lay User Profile}} \\ \hline
\textbf{Domain} & \textbf{Lay User ID} & \textbf{\begin{tabular}[c]{@{}c@{}}LLM \\ Familiarity\end{tabular}} & \textbf{Degree} & \textbf{Gender} & \textbf{Type of Assistance Received} \\ \hline
\multirow{5}{*}{Nutrition} & NutUsr1 & No & Bachelor's Degree & F & 1 on 1 Nutritional Support \\ \cline{2-6} 
                           & NutUsr2 & Yes & PhD & M & Online Nutrition Services \\ \cline{2-6} 
                           & NutUsr3 & No & Highschool Diploma & M & Guidance on Disease Management \\ \cline{2-6} 
                           & NutUsr4 & No & Master's Degree & F & Online Nutrition Services \\ \cline{2-6} 
                           & NutUsr5 & No & Master's Degree & F & 1 on 1 Nutritional Support \\ \hline
\multirow{5}{*}{Pedagogy}  & PedUsr1 & Yes & Bachelor's Degree & F & 1 on 1 Learning Support\\ \cline{2-6} 
                           & PedUsr2 & No & Bachelor's Degree & F & 1 on 1 Learning Support \\ \cline{2-6} 
                           & PedUsr3 & Yes & Master's Degree & M & 1 on 1 Learning Support \\ \cline{2-6} 
                           & PedUsr4 & No & Bachelor's Degree & F & Help from Teaching Assistant \\ \cline{2-6} 
                           & PedUsr5 & Yes & Master's Degree & M & 1 on 1 Learning Support \\ \hline
\end{tabular}
\caption{\revision{Lay User Participant Profile.}}
\Description{%
A table summarizing demographic and background information for lay user participants across two domains: Nutrition and Math Pedagogy.
Each row corresponds to a unique participant identifier.
Columns include domain, lay user ID, self-reported familiarity with large language models, highest degree attained, gender, and the type of educational or nutritional assistance previously received.
The table shows variation in participants’ educational backgrounds, prior exposure to LLMs, and support experiences within each domain.
}
\label{tab:layuser}
\end{table*}

\subsection{Scenario Creation and Output Generation}
\revision{Our case study uses a scenario-specific evaluation focused on a set of complex, domain-specific prompt instructions. This approach reflects a common practice in building real-world LLM applications, where teams iteratively test and refine a single core system prompt for a specific task before deployment~\cite{szymanski2024integrating, yang2024generating}. In these settings, the prompt’s contextual details determine what constitutes an appropriate or high-quality response. Therefore, studying prompt-specific criteria creation provides insight into how real-world developers and evaluators reason about quality in context.}

We crafted three prompt scenarios for each domain, nutrition and math pedagogy, based on a review of relevant literature. For the nutrition domain, prompt scenarios focused on disease management, food product suitability, and meal planning. Because dietitians are trained to approach diverse client needs, these prompt scenarios were selected to cover varied but common challenges that reflect typical areas of practice for dietitians and that LLMs could also potentially address~\cite{szymanski2024integrating, kirk2023comparison, ponzo2024chatgpt, garcia2023chatgpt}.
The pedagogy domain focused on math instruction, specifically algebra, as a foundation for understanding how evaluation criteria might be set in education. Prompt scenarios were inspired by existing studies on LLMs-as-tutors for secondary education and focused on common teaching methods, such as generating hints for students with partially completed solutions, providing feedback on student responses, and explaining concepts \cite{lieb2024student}. To create the prompt scenarios, we applied these teaching methods to standard algebra questions from the 9th-12th grade curriculum. Although math-focused, these tasks reflect pedagogical techniques that are widely applicable across other content areas requiring step-by-step instruction, as similarly seen in science or language learning. Detailed descriptions of each prompt scenario can be found in Appendix \ref{appendix:prompts}.

The outputs presented to participants were generated from the prompt scenarios using three different LLMs: GPT-4o, Claude 3.5 Sonnet, and Gemini 1.5 Flash. These models were selected to provide a range of responses to ensure that the criteria set by participants could be applied to variations seen across different LLMs.

\subsection{Study Protocol}
We conducted interviews with both domain experts and lay user participants to create and refine evaluation criteria. Domain experts were provided with prompt scenarios related to their specific fields, while lay users were presented with scenarios in nutrition or math pedagogy that mirrored the kinds of services they had previously received. The interviews lasted 1.5 hours and were conducted over Zoom. Before starting the study, we explained to participants that evaluation criteria should be created to serve as `rules' or `guidelines' to standardize information that they would expect in an output, and also to be used to evaluate the quality of outputs. 
Participants independently generated evaluation criteria in two separate phases:

\begin{enumerate}
    \item \textbf{Initial Criteria Creation (\textit{a priori}):} Participants were first presented with \revision{an LLM} prompt scenario and asked to read it over. We asked the participant to think about how they would like the prompt to be answered. We asked all the participants questions such as \textit{``What specific type of information should be included?''} The participants then created evaluation criteria that we labeled as \textit{a priori}.
    \item \textbf{Criteria Refinement (\textit{a posteriori}):} After developing the \textit{a priori} criteria, participants were presented with three different LLM-generated outputs related to the prompt scenario. They reviewed each output and considered whether or not their \textit{a priori} criteria applied. The participants noted aspects of the outputs that they liked, disliked, or would change and were asked to either refine their \textit{a priori} criteria or add additional criteria. We labeled all refined or added criteria as \textit{a posteriori}. 
\end{enumerate}

\begin{figure*}[!t]
\centering
\includegraphics[width=\textwidth]{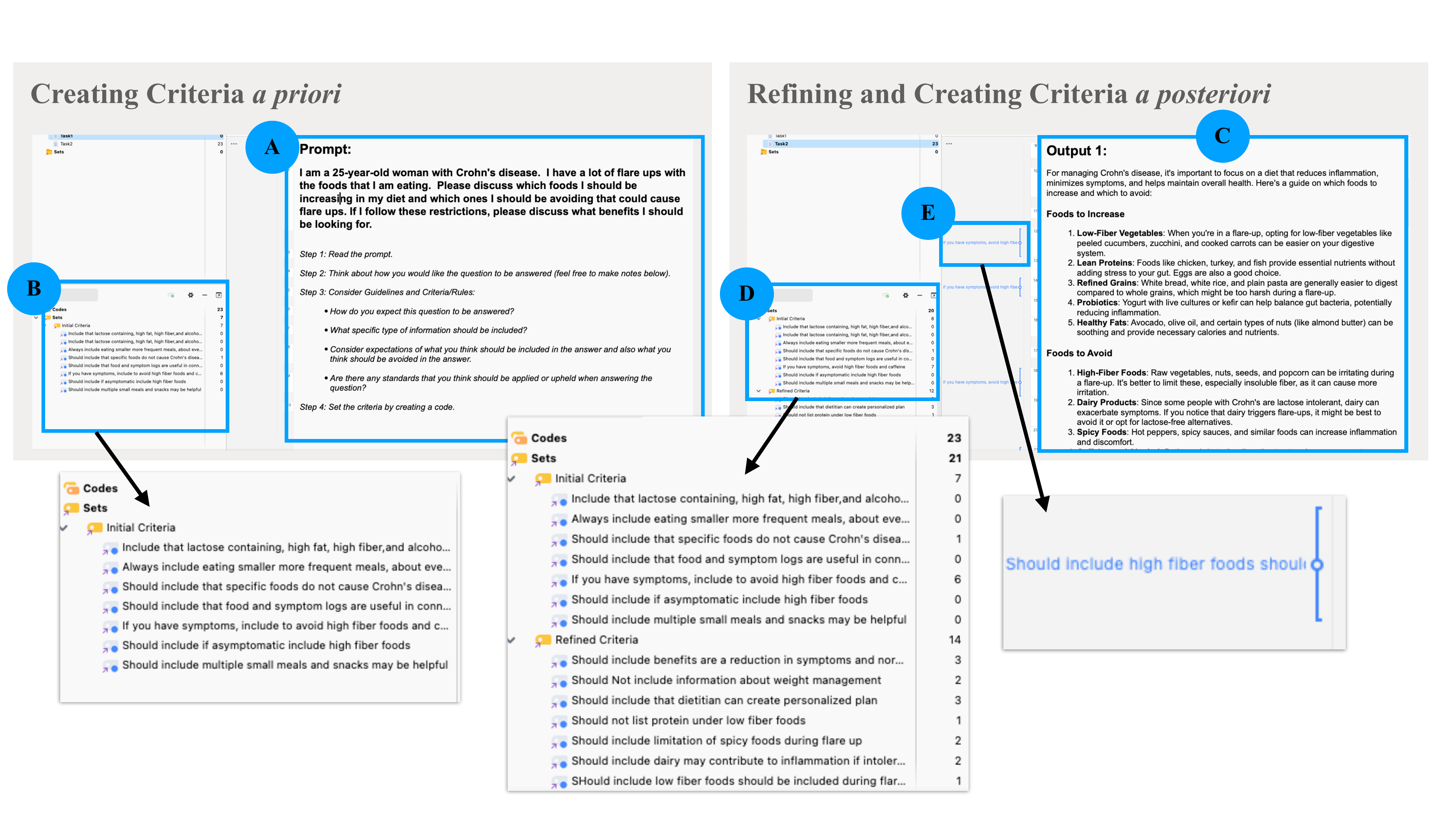}
\caption{\textbf{Workflow for Creating and Refining Evaluation Criteria in MAXQDA.} The left panel shows the \textit{a priori} phase, where participants (A) review a scenario-specific prompt and (B) create and store initial evaluation criteria. The right panel highlights the \textit{a posteriori} phase, where participants (C) review outputs generated by the LLMs, (D) either refine or create additional criteria, and (E) tag criteria to parts of the output during the \textit{a posteriori} phase.}
\label{fig:maxqdafigure}
\Description{The diagram illustrates the process for developing evaluation criteria in two phases. The left panel shows the a priori phase, where participants (A) review a scenario-specific prompt and (B) create initial evaluation criteria using a structured interface. The right panel highlights the a posteriori phase, where participants (C) review outputs generated by a large language model and (D) refine their criteria based on the outputs' content. Annotations (E) emphasize the interface elements used to facilitate the iterative refinement process. This workflow aims to capture participants' thought processes in setting and adapting evaluation criteria.}  
\end{figure*}

\revision{These LLM prompt scenarios and generated outputs were used in the evaluation process to allow participants to directly consider the constraints and expected behavior of the LLM.} Each participant evaluated the same set of 3 scenarios and 3 corresponding outputs to allow for a comparative analysis of criteria development across participants.
To mitigate potential learning effects and order bias, we counterbalanced the prompt scenarios across participants. This means that each participant received the scenarios in a different order to ensure that no single scenario was consistently presented first. Additionally, the LLM outputs were randomized to prevent any bias associated with the sequence of outputs. Participants typically spent approximately 10 minutes in the \textit{a priori} phase and approximately 15 minutes in the \textit{a posteriori} phase, totaling approximately 25 minutes evaluating each of the three scenarios.

\subsubsection{Facilitation of Criteria Development} For the development and tracking of evaluation criteria, we utilized MAXQDA, a software tool designed for qualitative data analysis, specifically suited for coding and tagging outputs based on user-defined categories \cite{marjaei2019maxqda}. Figure \ref{fig:maxqdafigure} describes the system and features using the MAXQDA system during the study. MAXQDA allowed participants to independently create system codes representing their criteria. In addition to allowing for an easy way to create initial \textit{a priori} criteria, MAXQDA provided an easy mechanism for refinement. During the \textit{a posteriori} phase, participants could either revise or add new criteria as they reviewed the outputs. The tool allowed participants to ``tag'' outputs with both their original or newly refined criteria, making it possible to iteratively update and expand their evaluations based on what they observed in the LLM outputs. This systematic tagging and refinement process ensured that participants could effectively track how the criteria applied across different outputs and make necessary adjustments to capture relevant evaluation measures.

\subsection{LLM Generation of Criteria} 
To parallel the process followed by human participants, we used five different types of commonly referenced LLMs to create the evaluation criteria: GPT-4o (LLM1), GPT-4 (LLM2), gemini-1.5-flash (LLM3), claude-3-5-sonnet-20241022 (LLM4), and claude-3-5-haiku-20241022 (LLM5)~\cite{cheng2023now, king2023introduction, liu2023chatcounselor}. The LLMs were tasked with generating evaluation criteria in two phases. In the first phase \textit{(a priori}), each LLM was provided with the same prompt scenario given to human participants and was instructed to generate an initial set of criteria based solely on the prompt. 
In the second phase \textit{(a posteriori}), each LLM was prompted with its initial criteria generated in the first phase, along with the same three LLM-generated outputs shown to the human participants.  
Each LLM was then instructed to refine its initial criteria or develop new ones based on its evaluation of these outputs. 

\revision{The prompt instructions were intentionally minimal and designed to elicit single-step, baseline behavior. Consistent with rubric-generation patterns used in state-of-the-art LLM-as-a-Judge frameworks such as EvalLM~\cite{kim2024evallm}, no specialized prompting strategies (e.g., Chain-of-Thought, multi-step workflows) were applied. This approach mirrored the task framing given to human participants and was chosen to fulfill our core objective, which was to enable a controlled and comparable evaluation across domain experts, lay users, and LLMs.} A detailed description of the system instructions used to generate the criteria can be found in Appendix \ref{appendix:criteria_generation}.



\subsection{Criteria Consolidation and Analysis} 
We followed standard open-coding procedures \cite{williams2019art} to analyze the generated criteria from the three different groups (domain experts, lay users, and LLMs). Three researchers independently participated in the coding process, beginning by thoroughly reviewing the criteria from the interviews to gain familiarity with the data. They then created initial codes and coded 1---2 interviews each based on their findings. They then reconvened in meetings to discuss and reconcile these initial codes, which we refer to as sub-themes. Subsequently, the rest of the interviews were coded using these codes. New codes were added as they emerged from the interviews. These sub-themes varied by scenario, reflecting the unique context and focus of each task. After developing the initial sub-themes, they were then organized into higher-level themes that were consistent across all three scenarios and both domains. These themes were consolidated and evolved into key findings, which are detailed in Section \ref{results}.

The criteria were also labeled according to whether they were created or modified either during the initial prompt-only phase \textit{(a priori)} or after the participants reviewed the outputs \textit{(a posteriori)}. This temporal consideration allowed us to analyze how the criteria evolved over time and to compare the themes developed by each group across the two phases of evaluation. 
In addition, for each scenario, source, and domain, we calculated the average number of criteria created in each phase.

\section{Findings}
\label{results}

In this section, we present our findings for RQ1, which examines the similarities and differences of criteria created by domain experts, lay users, and LLMs, as well as for RQ2, which compares the types of criteria and themes across \textit{a priori} and \textit{a posteriori} phases. 
We have organized our analysis into key findings (KF).

\subsection{Similarities and Differences among Domain Experts, Lay Users, and LLMs (RQ1)} \label{sec:RQ1_findings}

Notable differences in criteria development among the three groups are presented in the key findings below. Table \ref{tab:evaluation_criteria_examples} provides examples of the types of criteria among the three groups.\\

\noindent
\textbf{KF1: Domain experts create knowledge-informed and contextually grounded criteria that extend beyond the content in the prompt scenario.}

Domain experts drew upon their specialized knowledge to establish criteria that reflected professional principles and standards rather than simply echoing the details of the prompt. This behavior was especially prominent in the \textit{a priori} phase, where experts formulated criteria immediately after reading the prompt scenario and before viewing a model-generated response. Their criteria reflected proactive disciplinary expertise combined with an orientation toward user needs.

In the nutrition domain, the term \textit{knowledge-informed} means drawing directly on evidence-based dietary science. For example, when presented with a scenario for breakfast recommendations, dietitians explicitly created criteria that specified the precise thresholds, such as minimum gram ranges of protein to sustain satiety, limits on added sugar relative to daily value \% (DV), and specific targets for fiber intake based on clinical guidelines. In a scenario around weight loss, dietitians calculated and included acceptable calorie deficits and estimated the resulting daily or weekly weight loss, including criteria on specific diet plans to avoid. These criteria demonstrate that experts drew on disciplinary knowledge and behavior change strategies to create scientifically grounded evaluation criteria that extended well beyond the prompt.

In the math pedagogy domain, evaluation criteria were \textit{knowledge-informed} through grounding in disciplinary training in mathematics education. Rather than echoing the prompt, they translated scenarios into concrete instructional requirements rooted in pedagogical best practices. For example, in the ``providing hints'' scenario, experts required outputs to articulate specific solution strategies, such as applying the Order of Operations or isolating variables step by step, while also suggesting alternate paths forward. In the ``feedback on errors'' scenario, they specified that outputs should explicitly identify the student’s mistake (e.g., dividing when they should have subtracted) and frame it in a way that guides re-teaching. Finally, in the ``explaining concepts'' scenario, experts required inclusion of the precise mathematical representation (the standard quadratic form $ax^{2} + bx + c$) along with a rationale for its use. These examples illustrate how experts drew directly on disciplinary knowledge to establish evaluation criteria that emphasized conceptual accuracy, instructional clarity, and pedagogical utility beyond simply restating the prompt’s directions.

Domain experts also created criteria that were \textit{contextually grounded}, meaning that experts continually situated their evaluative criteria in relation to the imagined user’s needs. For example, dietitians emphasized creating criteria to prompt for a person’s dietary history, mental state, and preferences, recognizing that a technically ``correct'' recommendation might fail if it overlooked a user’s motivation or past experiences. Similarly, in math pedagogy, educators did not only require correct explanations of formulas such as the quadratic equation; they insisted on contextually appropriate instructions, such as breaking problems into manageable steps, providing problem examples, or probing student understanding with follow-up questions.\\

\begin{table*}[t]
\centering
\caption{Evaluation Criteria Examples Across Domains and Participant Types}
\Description{%
A multi-row table comparing example evaluation criteria across two domains: Nutrition and Math Pedagogy. 
Rows are grouped by domain and scenario, while columns show example criteria written by Domain Experts, Lay Users, and Large Language Models.
Nutrition scenarios include breakfast recommendations, Crohn’s disease dietary guidance, and weight loss advice.
Pedagogy scenarios include providing hints, feedback on errors, and explaining mathematical concepts.
The table illustrates differences in focus across participant types, such as experts emphasizing safety and correctness, lay users emphasizing clarity and explanation, and LLMs emphasizing comprehensive or procedural guidance.
}
\label{tab:evaluation_criteria_examples}
\renewcommand{\arraystretch}{1.4}
\rowcolors{3}{gray!10}{white}

\begin{tabular}{%
  >{\raggedright\arraybackslash}p{1.5cm}
  >{\raggedright\arraybackslash}p{2.5cm}
  >{\raggedright\arraybackslash}p{4.0cm}
  >{\raggedright\arraybackslash}p{4.0cm}
  >{\raggedright\arraybackslash}p{4.0cm}
}
\toprule
\textbf{Domain} & \textbf{Scenario} & \textbf{Domain Expert} & \textbf{Lay User} & \textbf{LLMs} \\
\midrule

Nutrition & 1: Breakfast Recommendations &
Recommend cereals with less than 20\% of the DV for added sugars, at least 10\% DV for fiber, and at least 10\% DV for protein. (NutExp4) &
Detail every ingredient in this cereal and detail all of the nutrition facts like protein \% or fat \%. (NutUsr1) &
Explanation includes a detailed assessment and potential impact of added sugars in the cereal. (LLM1) \\

& 2: Crohn's Disease &
Should tell person (to prevent excess fluid) that fluid intake goals are the person's body weight in pounds divided by 2 to get how many ounces of fluids. (NutExp2) &
Output should include basic definition/impact of having Crohn's disease. (NutUsr4) &
The response offers clear, actionable dietary advice tailored to a 25-year-old woman with Crohn's disease. (LLM4) \\

& 3: Weight Loss &
Should not include specific diets such as ketogenic or intermittent fasting. (NutExp3) &
Should contain examples of high protein, low sugar food options. (NutUsr5) &
The response advises against any potentially harmful or unsustainable diet plans. (LLM3) \\

\midrule

Pedagogy & 1: Providing Hints &
Provide alternate approaches such as Order of Operations which allows for common denominators to be sought out immediately or after subtracting 5 from both sides. (PedExp3) &
Walk them through the solving pipelines from the beginning, and point out where they are. (PedUsr3) &
The response provides guidance without directly solving the entire problem for the student. (LLM5) \\

& 2: Feedback on Errors &
State that student had an error in solving for t. It appears you divided when you should have subtracted. (PedExp5) &
If there is anything wrong, tell user and prompt with questions to re-teach. (PedUsr1) &
The response includes appropriate mathematical notation and step-by-step calculations. (LLM4) \\

& 3: Explaining Concepts &
Should include the standard form of the equation (i.e., \( ax^2 + bx + c \)). (PedExp4) &
A brief intro to quadratic equations (a few sentences) as a preface to ease the user into understanding the concept/need. (PedUsr2) &
The response correctly states the quadratic formula. (LLM1) \\

\bottomrule
\end{tabular}
\end{table*}

\noindent
\textbf{KF2: Lay users develop criteria that reflect their personal needs and place greater emphasis on understandability, layout, and format compared to domain experts and LLMs.}

Lay users approached creating evaluation criteria not through technical standards or professional guidelines, but through the lens of everyday \textit{practicality} and \textit{accessibility}. Rather than incorporating technical standards, their criteria reflected immediate concerns such as whether ingredient information was clearly listed, whether the food recommendations were cost-effective or within their budget, or whether math steps were explained in a simple and easy-to-follow language. Across both domains, lay users frequently emphasized the structure and delivery of information, introducing criteria such as using bullet points, clear headings, or visual aids to ensure that outputs be organized and accessible. 

In nutrition, this often meant expecting outputs that clearly summarized nutritional facts, presented ingredients in a transparent way, and communicated dietary advice in plain, non-technical language to aid everyday food choices. In math pedagogy, this translated into criteria that prioritized step-by-step walkthroughs of problems, simple and direct phrasing, and supports such as visual representations to make abstract processes easier to follow. While their contributions were less technical, they focused on important usability and comprehension factors that influence how end users engage with information. Some lay users also suggested criteria that reflected preferences for how learning should be supported, including help with recognizing errors or encouraging logical sequencing, which offered practical insights that may not be captured by experts or LLMs.\\


\noindent
\textbf{KF3: The LLMs create criteria that are more generic and limited to the key aspects of the prompt scenario relative to the other groups.}

LLMs generated evaluation criteria that \textit{closely mirrored
the language and intent of the prompt}. While criteria often touched on many of the same thematic areas, those generated by LLMs often lacked the depth, contextual awareness, and reasoning observed in domain experts and, at times, even lay users. For instance, in the nutrition domain, the LLM might suggest that a response should include cereal substitutions or offer dietary advice for Crohn's disease, without specifying the nutritional benchmarks or behavioral strategies that experts would include. In math pedagogy, LLMs produced broad criteria such as explaining misconceptions or avoiding errors, which, while relevant, did not capture the instructional depth or learner-centered perspectives introduced by human participants. While the LLMs occasionally introduced features such as aligning explanations with a student’s educational level or promoting a supportive tone in math tasks, these remained surface-level. Overall, the criteria created by LLMs reflected a fairly generic grasp of the prompt's intent but rarely demonstrated the complex decision-making or adaptive reasoning shown by human participants.

\begin{figure*}[t]
    \centering
    \begin{subfigure}[t]{\textwidth}
        \centering
        \includegraphics[height=6cm]{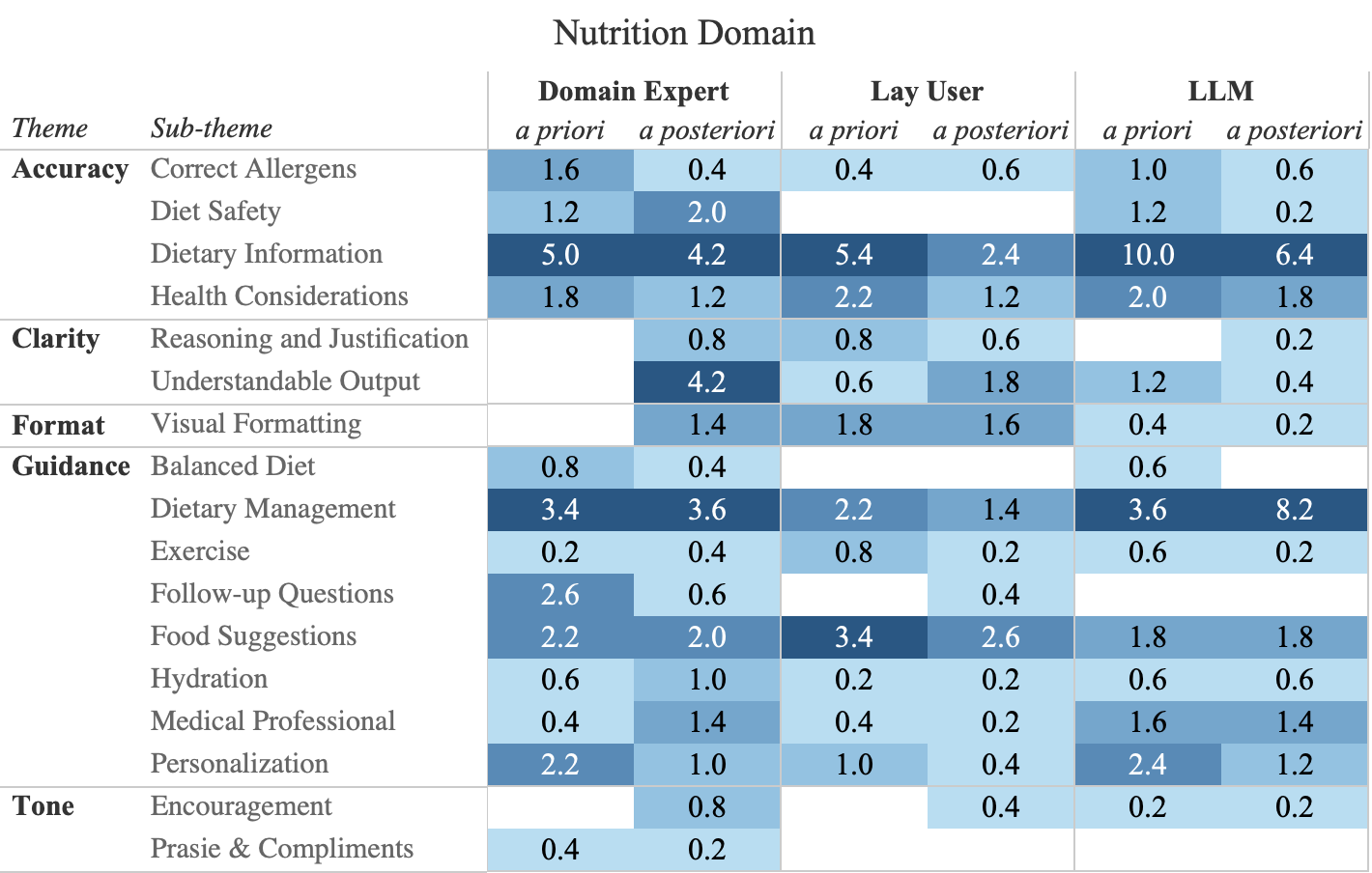}
        \caption{\revision{Average Number of Criteria Created for the Nutrition Domain.}}
        \label{sub_fig:heatmap_nutrition}
    \end{subfigure}

    \vspace{0.5cm} 

    \begin{subfigure}[t]{\textwidth}
        \centering
        \includegraphics[height=6cm]{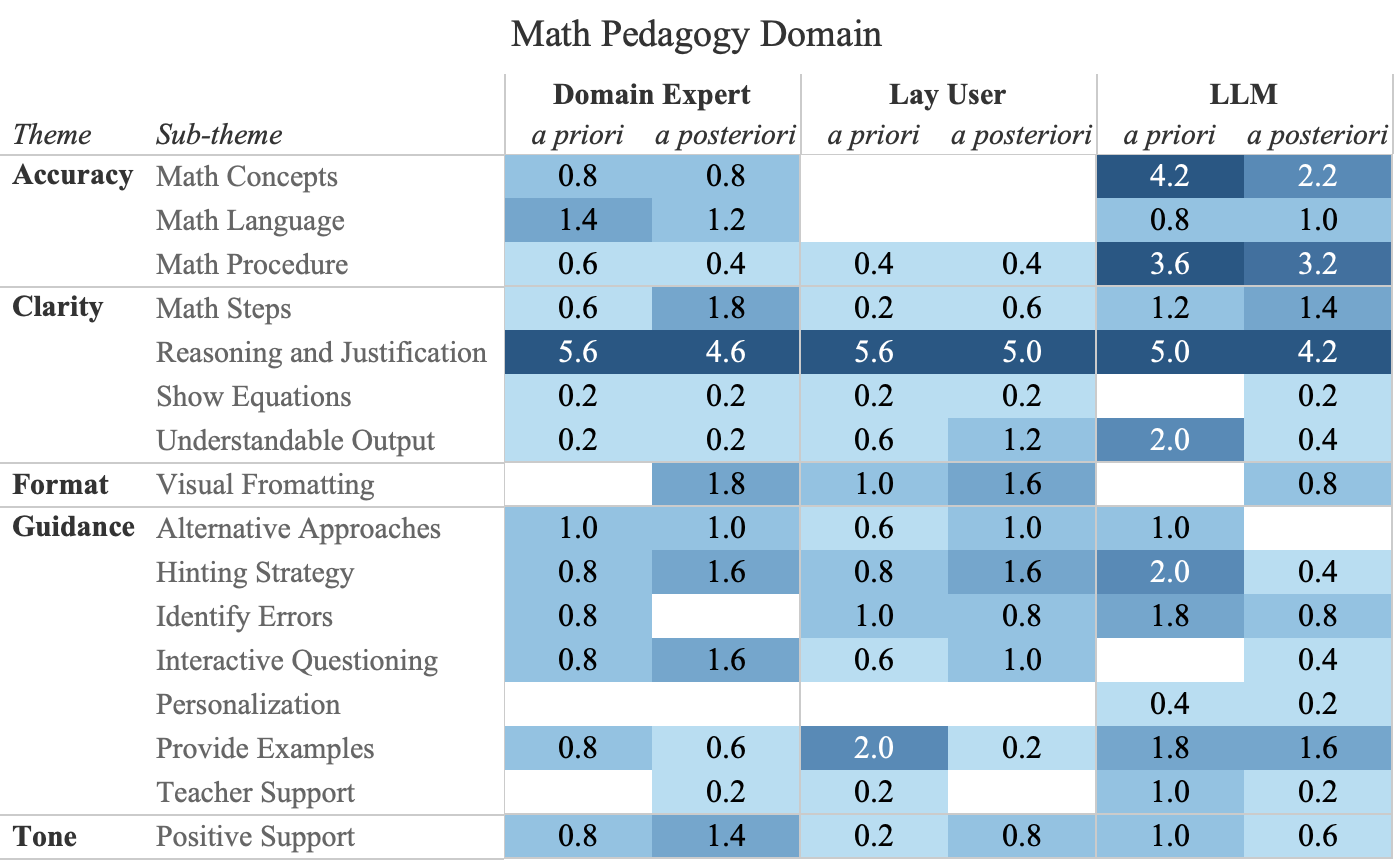}
        \caption{\revision{Average Number of Criteria Created for the Math Pedagogy Domain.}}
        \label{sub_fig:heatmap_pedagogy}
    \end{subfigure}

    \caption{\revision{Heat map of the average number of criteria for each domain created by experts, lay users, and the LLMs at both the \textit{a priori} and \textit{a posteriori} phases.}}
    \Description{Heatmap tables showing the average number of evaluation criteria created across Nutrition and Math Pedagogy domains. Rows represent evaluation themes and sub-themes, columns compare Domain Experts, Lay Users, and LLMs at a priori and a posteriori stages. Darker blue indicates higher average counts. Across both domains, Domain Experts emphasize accuracy and reasoning, Lay Users emphasize clarity and understandability, and LLMs emphasize procedural and content coverage.}

    \label{fig:heatmap}
\end{figure*}

\subsection{Criteria Evolution Over Time (RQ2)}
In this section, we will discuss how the evaluation criteria differ among groups between the \textit{a priori} and \textit{a posteriori} phases. Figure \ref{fig:heatmap} presents a heatmap showing the average number of criteria created by each group across sub-themes within the nutrition and pedagogy domains at both stages of evaluation.
\newline

\begin{figure*}[t]
    \centering

    \includegraphics[width=0.65\textwidth]{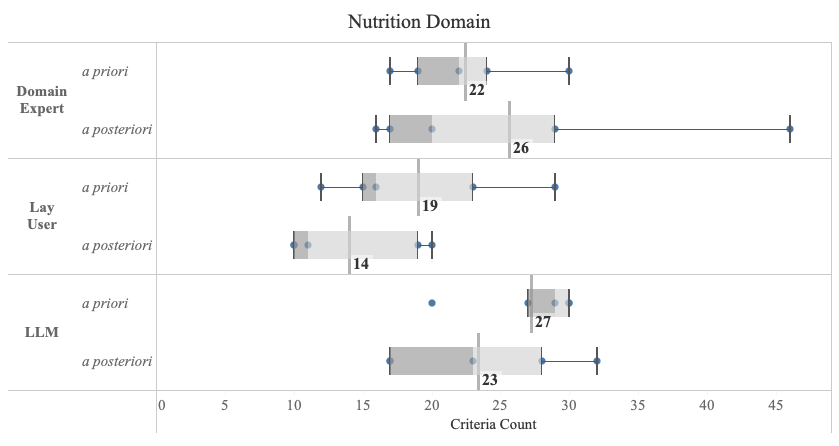}
    \vspace{0.75em}

    \includegraphics[width=0.65\textwidth]{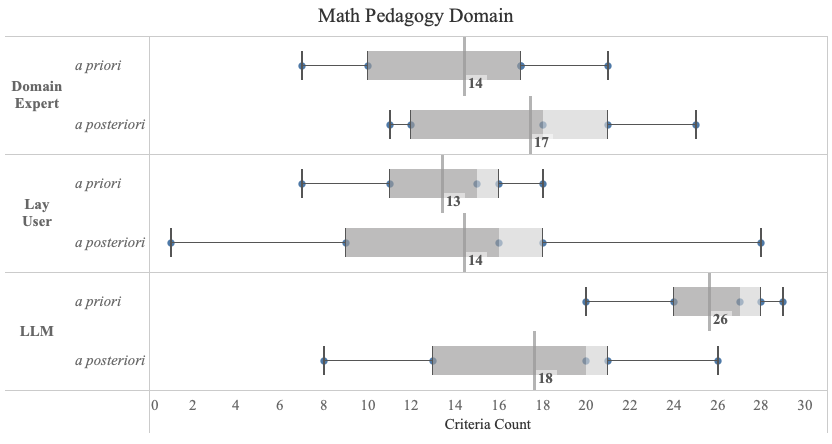}

    \caption{\revision{Distribution of criteria counts generated in the \textit{a priori} and \textit{a posteriori} phases across participant types (Domain Experts, Lay Users, and LLMs) in the Nutrition and Math Pedagogy domains. Means are labeled within each distribution.}}
    \Description{The figure shows two horizontal plots—one for the Nutrition domain and one for Math Pedagogy—each displaying criteria counts for Domain Experts, Lay Users, and LLMs in the a priori and a posteriori phases. For each group and phase, a light gray box shows the distribution of counts, whiskers mark the range, and a point with a numeric label marks the mean. In both domains, Domain Experts show a slight increase in criteria counts after reviewing outputs, Lay Users show small decreases or modest increases depending on the domain, and LLMs consistently show lower counts in the a posteriori phase. The figure highlights how the criteria generation shifts once participants see concrete outputs.}
    \label{fig:boxplots}
\end{figure*}

\noindent
\textbf{\revision{KF4: Across both domains, \textit{a posteriori} criteria counts remained close to \textit{a priori} levels for all groups.}\delete{In the \textit{a posteriori} phase, an increase in the number of evaluation criteria developed relative to the \textit{a priori} phase.}}

\delete{As shown in Table~\ref{tab:num_table}}\revision{Figure \ref{fig:boxplots} shows that, across both domains, the number of criteria generated in the \textit{a priori} and \textit{a posteriori} phases remained relatively similar. Domain experts produced slightly more criteria on average in the \textit{a posteriori} phase, suggesting a tendency to expand their evaluations after reviewing concrete outputs. In contrast, LLMs generated fewer criteria on average in the \textit{a posteriori} phase. Overall, the key takeaway is that all sources continued to produce a substantial number of criteria even after viewing the outputs, indicating consistent engagement across phases.} \delete{in both domains, experts generated more criteria during the \textit{a posteriori} phase compared to the \textit{a priori} phase. In contrast, both lay users and LLMs created fewer criteria in the \textit{a posteriori} phase.} We also noted that domain experts, lay users, and LLMs in the nutrition domain created \revision{on average} more criteria in both phases compared to those in pedagogy. This may suggest that tasks in the nutrition domain prompt a greater need for detailed and specific criteria, likely due to the complex nature of health-related decisions, compared to the more structured and instructional nature of the pedagogy domain. \delete{As shown in Figure~\ref{fig:boxplots}, in both domains, experts generated more criteria during the \textit{a posteriori} phase compared to the \textit{a priori} phase. In contrast, both lay users and LLMs created fewer criteria in the \textit{a posteriori} phase.}\\

\noindent
\textbf{KF5: In the \textit{a posteriori} phase, domain experts and lay users frequently introduced new criteria in response to missing, incorrect, or insufficient information in the outputs.}

We observed that domain experts and lay users developed new evaluation criteria after reviewing the output, specifically in response to gaps, errors, or areas that they felt required improvement. As previously noted in KF3, the LLM-generated criteria tended to reflect information already present in the output and did not include new criteria addressing missing or problematic content.

Domain experts introduced new criteria when they encountered vague or misleading information. For instance, one expert responded to unclear phrasing by adding the criterion: \textit{``Remove opinionated text''} (NutExp5), while another revised imprecise nutritional language by specifying: \textit{``State `high' not `likely high' if DV [Daily Value \%] for added sugar is >20\%''} (NutExp4). In a scenario involving weight loss, an expert added a new criterion to explicitly state which diets to exclude: \textit{``Should not include paleo diet; should not include carb cycling''} (NutExp3), which was prompted by an output that recommended these potentially harmful diets, which the LLM failed to flag. In the education domain, experts added criteria when outputs lacked specificity or included unnecessary detail. For example, after encountering an overly descriptive explanation of the discriminant, one expert added: \textit{``Should not describe what the discriminant does''} (PedExp2). Others added criteria to require clearer explanation steps, such as: \textit{``Give more detail: 5 = 5/1 = showing that you multiply the numerator and denominator by 2''}, or broader coverage such as: \textit{``Include additional considerations: domain restrictions, discriminant implications, negative vs. positive coefficients''} (PedExp3). These examples illustrate that domain experts continued to apply their professional knowledge in the \textit{a posteriori} phase to generate new criteria that refine, extend, and correct what was observed in the outputs.

Lay users also introduced new criteria in response to omissions or presentation issues, though their focus often remained on clarity and accessibility. In the nutrition domain, they requested clearer framing around dietary goals, such as: \textit{``Include more explanation on which [diets] would help lose weight vs. which ones would help build muscle''} (NutUsr1), \textit{``Specify calorie threshold''} (NutUsr3), and \textit{``Provide information on maintenance calories''} (NutUsr2). They also reacted positively to the inclusion of actionable advice, adding criteria such as: \textit{``Output includes tips for how to implement diet (tips for success)''} (NutUsr4). In the pedagogy domain, lay users critiqued the instructional framing and added criteria such as: \textit{``Should prompt user and guide them to the correct answer rather than showing error right away''} (PedUsr2) and \textit{``Should leave the student to finish the rest of the problem''} (PedUsr4). These examples reflect a key distinction in how lay users approached evaluation: while domain experts focused on \emph{what} information should be provided, lay users emphasized \emph{how} the information should be presented to support their understanding and decision-making.

Notably, there were also instances where LLMs and domain experts contradicted each other in their newly introduced criteria in the \textit{a posteriori} phase. In one nutrition scenario, the LLM added the criterion: \textit{``The response explains the relationship between refined grains and glycemic index'' (LLM4)}, while the expert added: \textit{``Don’t provide the term glycemic index'' (NutExp2)}. In the pedagogy domain, experts preferred that outputs stop after providing a first-step hint, \textit{``Response should stop at step 1. Other steps are not needed until step one is done'' (PedExp5)}, while the LLM, in contrast, added the criterion: \textit{``The response offers a step-by-step breakdown of the next steps in solving the equation''} (LLM1), which emphasized completeness over instructional pacing.

This reactive and reflective development of criteria demonstrates a key difference in how human evaluators and LLMs create evaluation criteria in the \textit{a posteriori} phase. While LLMs responded more descriptively to what was visible, human participants, especially domain experts, exhibited a more critical and constructive stance. Their criteria functioned not only as assessments but as correctives, actively shaping what a high-quality output should look like.\\

\noindent
\textbf{KF6: In the \textit{a posteriori} phase, evaluation criteria began to converge across domain experts, lay users, and LLMs, often reflecting features that were explicitly present in the output.}

Alongside the development of new criteria in response to missing or insufficient information (as discussed in KF5), we also observed a distinct pattern of convergence in the \textit{a posteriori} phase: participants across all groups developed criteria around features already present in the generated outputs. This alignment appeared to result from a shared recognition of useful or well-structured content, rather than deeper contextual reasoning. 

The first area of convergence was observed in the themes of clarity and guidance. For example, in the nutrition domain, one output stated that the product was a poor dietary choice for the user and why, at the beginning of the output. In response, experts created criteria that reflected and reinforced this structure, including: \textit{``Should include if it is an ideal choice right away''} (NutExp2) and \textit{``For each diet suggested, should include focus and reason why''} (NutExp3). Similarly, in the math pedagogy domain, an output presented a worked example of solving a quadratic equation with a clear step-by-step breakdown. In response, one expert added the criterion: \textit{``Breaks down standard form of a quadratic equation''} (PedExp4), while the LLM generated: \textit{``The response provides a step-by-step example of using the quadratic formula to solve a quadratic equation''} (LLM4). This illustrates how, in the \textit{a posteriori} phase, both human participants and LLMs tended to build criteria around what was already articulated in the model output.

A separate area of convergence occurred around tone and visual presentation. In both domains, participants from all groups reacted positively to outputs that used clear structure, headings, bullet points, or lists by developing criteria that reflected the value of clear visual organization. While lay users frequently prioritized these elements \textit{a priori}, domain experts and LLMs began to echo these preferences after encountering differences in the formatted outputs. For example, one domain expert added: \textit{``Should include a bulleted list of foods beneath recommendations''} (NutExp3), while an LLM noted: \textit{``The response is well-organized and logically structured, making it easy for the user to follow''} (LLM3). Similarly, while domain experts initially prioritized instructional or clinical precision in the \textit{a priori} phase, they later added criteria for demonstrating encouragement and positive support in the \textit{a posteriori} phase, influenced by outputs that effectively modeled those characteristics. For example, in the nutrition domain, an expert added the criterion: \textit{``Always encourage dietary changes that can be maintained long term''} (NutExp2), and in the pedagogy domain, another expert suggested: \textit{``Should affirm correct student thinking''} (PedExp2), both directly responding to positive tone and affirming language present in the respective outputs.

Finally, we observed that LLMs occasionally adopted criteria that were originally introduced by domain experts in the \textit{a priori} phase, but only developed these criteria after encountering related features in the output. In the nutrition domain, for example, while a domain expert created a criterion \textit{a priori} requiring that sugar content be explicitly stated in cereal recommendations, the LLM only introduced a similar criterion after observing this detail in one of the outputs, suggesting: \textit{``The response clearly identifies the high sugar content in Great Value Cinnamon Crunch Breakfast Cereal as problematic for blood sugar management''} (LLM5). Likewise, in the education domain, domain experts had developed criteria in the \textit{a priori} phase, evaluating the inclusion of specific equations or step-by-step breakdowns, which were later echoed by both lay users and LLMs after these features appeared in one of the model's responses. For instance, the LLM added: \textit{``The response identifies that multiplying both sides by 4 is the next logical step to eliminate fractions''} (LLM4), reflecting how the outputs solved the equation.
\\

\section{Discussion}


This paper contributes new insights into the strengths and weaknesses of domain experts, lay users, and LLMs in creating evaluation criteria for domain-specific tasks\revision{, and suggests that their complementary strengths can be leveraged to improve evaluation workflows.} \delete{Our findings suggest that we can apply the complementary strengths among the three sources to improve evaluation workflows.}
In this section, we will discuss our findings \delete{on the similarities and differences in evaluation criteria} among the three sources and two evaluation phases (RQ1 and RQ2) and suggest design implications (RQ3) for evaluation workflows to ensure the outputs meet the standards of both domain experts and lay users.  

\subsection{Understanding Where and How Perspectives Align in Evaluation}
While prior research has emphasized the importance of user-centered evaluation frameworks and has explored the potential of LLMs as customizable evaluators~\cite{pan2024human}, our findings extend this work by demonstrating the critical role of multiple perspectives in the process.
Our study reveals the strengths that each group contributes to evaluations and the stages at which they are most impactful. As evaluation workflows are developed, these findings point to how diverse perspectives can be strategically incorporated, as well as where there is alignment across groups, to effectively guide the design of more comprehensive and balanced evaluation processes.




As has been shown in previous studies, integrating expert knowledge into LLMs leads to more accurate and reliable outputs~\cite{szymanski2024integrating, zhang2024way}. Our study complements these results, by showing that domain experts provide specific and targeted criteria that address both knowledge-based aspects from the prompt and also consider the long-term outcomes for the end-user. 
Expert evaluation criteria align with professional standards and are tailored to promote sustained improvements in user understanding and decision-making over time.
We observe that experts approach the task of creating criteria by answering the prompt as they would for a client or student, offering detailed guidance and outlining key information that should be included in the output, almost as if they are creating a checklist of what the response should contain. 
As discussed in KF5, domain experts frequently introduced criteria to improve or correct future outputs, which also supports prior findings that suggest ethical concerns related to bias and misinformation in LLMs can be mitigated by incorporating expert oversight~\cite{szymanski2024limitations, cheong2024not, Tokayev_2023}. Our work indicates that as HCI developers consider future evaluation workflows, the incorporation of experts should be considered to ensure the outputs are evaluated up to domain standards with the most educational impact for the user.

While expert criteria are more knowledge-based, the lay users approach the development of criteria based on \emph{what} they need to know and \emph{how} they want the information presented.
The lay users were observed to set their criteria with more of a focus on inquiry, and prioritizing clear and understandable presentations, often emphasizing formatting or visuals to enhance readability. 
These contributions are more short-term in impact, compared to those of the knowledge-driven criteria provided by domain experts.
In the short term, these criteria improve the user’s experience by enhancing the usability of the system and could lead to greater trust and engagement. 
This aligns with prior research on LLM interfaces where there were correlations in high levels of user trust based on ease of usability and how clearly the information was delivered~\cite{sun2024trust}. 
When developing evaluation workflows, it is crucial to incorporate both the immediate \delete{usability} concerns addressed by lay users and the deeper, more sustained contributions provided by domain experts.
The approach that we recommend differs from previous approaches to LLM evaluation, which have often focused only on developers or designers creating criteria and testing prompt outputs for alignment \cite{shankar2024validates, arawjo2024chainforge}.

While the LLMs occasionally produced criteria that matched themes from experts or lay users, the suggestions were often generic and narrowly focused on what was explicitly stated in the prompt or the output itself. 
\revision{Our results indicate that these patterns may arise from both the single-shot prompting strategy used in this study~\cite{zamfirescu2023johnny} as well as the inherent ways in which LLMs process information~\cite{ranjan2024comprehensive}.}
\revision{In the \textit{a priori} phase, prompting with the scenario alone (no output or domain knowledge) led to high-level criteria useful for basic requirement checks but lacked expert-level specificity.} 
\revision{Conversely, in the \textit{a posteriori} phase, where the output was included, the LLMs tended to derive criteria directly from the output, which risks reinforcing or legitimizing any errors or omissions present.}
\revision{These behaviors align with prior work showing that LLM-generated evaluations are highly sensitive to prompt design and the LLM's inherent tendency to anchor on the most immediate contextual cues~\cite{niu2025llama, wu2025answer}.} 
\revision{As LLMs are increasingly used to supplement human evaluators for cost-effectiveness~\cite{hamalainen2023evaluating, gilardi2023chatgpt}, it becomes essential to understand how developers can use LLMs responsibly for criteria generation tasks while mitigating risks of bias or misinformation.}
\revision{Although more structured prompting strategies could be used to improve LLMs and therefore could lead to different criteria generation~\cite{kim2023better}, such techniques cannot fully eliminate deeper challenges, as LLMs may still draw from outdated or contaminated data sources~\cite{szymanski2024integrating}, and may misrepresent demographic identities or reproduce societal biases~\cite{wang2024large}. These limitations suggest that while LLMs could be prompted to generate useful high-level criteria and support scalable evaluation workflows, they should not be used in isolation for tasks requiring domain-specific judgment or high-stakes evaluation.}
\delete{While we can utilize LLMs to produce evaluation criteria that ensure that the outputs address high-level prompt requirements, there are perceived risks we see with relying only on LLMs for this purpose. In the a posterior phase, LLMs often create criteria directly from the output, which means they may reinforce or legitimize inaccurate or misleading content if the output itself is flawed. In addition, LLMs often pull from outdated or contaminated data sources, which is particularly problematic in complex fields where inaccurate and potentially harmful information~\cite{szymanski2024integrating} can be spread. In addition, LLMs struggle to accurately represent human demographic identities and introduce biases that can be harmful in sensitive areas~\cite{wang2024large}. As LLMs are increasingly used to replace human participants for cost-effectiveness~\cite{hamalainen2023evaluating, gilardi2023chatgpt}, it is important to consider these harms before deploying LLMs for producing evaluation criteria. Our work demonstrates that while the LLM may be beneficial in producing high-level criteria to satisfy prompt requirements, it is important to incorporate diverse perspectives to prevent the dissemination of inaccurate or biased information.}

\begin{table*}[t]
\centering
\caption{Staged Evaluation Approach: Recommended Roles and Contributions}
\Description{%
A conceptual table outlining a staged evaluation workflow for large language model systems.
Rows correspond to evaluator types—domain experts, lay users, and LLMs—while columns describe when each evaluator should be involved, why they are included, their primary contributions, associated trade-offs, and related design guidelines.
The table emphasizes complementary roles: domain experts establish and validate domain-specific criteria, lay users focus on clarity and usability during output review, and LLMs support scalable, lower-level checks.
Overall, the table highlights ethical and practical trade-offs between expertise, cost, scalability, and risk across different stages of evaluation.
}
\label{tab:staged_evaluation}
\small
\renewcommand{\arraystretch}{1.4}

\begin{tabular}{%
  >{\raggedright\arraybackslash}p{1.4cm}
  >{\raggedright\arraybackslash}p{1.8cm}
  >{\raggedright\arraybackslash}p{3.6cm}
  >{\raggedright\arraybackslash}p{3.6cm}
  >{\raggedright\arraybackslash}p{3.6cm}
  >{\raggedright\arraybackslash}p{1.3cm}
}
\toprule
\textbf{Evaluator Type} & \textbf{When to Involve} & \textbf{Why} & \textbf{Primary Contributions} & \textbf{Trade-offs / Considerations} & \textbf{\revision{Design Guideline}} \\
\midrule

\vspace{-0.5\baselineskip}\textbf{Domain Experts} &
\begin{tightitemize}
  \item \textit{a priori} \newline \newline
  \item \textit{a posteriori} \newline (if feasible)
\end{tightitemize}
&
\begin{tightitemize}
  \item Establish detailed, domain-specific criteria aligned with professional standards.
  \item Review outputs to address omissions, inaccuracies, or contextual misalignments.
\end{tightitemize}
&
\begin{tightitemize}
  \item Develop knowledge-based criteria that reflect professional goals and standards.
  \item Create criteria informed by experience that extend beyond the content in the prompt scenario.
  \item Establish criteria to address missing or misstated content.
\end{tightitemize}
&
\begin{tightitemize}
  \item Setting detailed \textit{a priori} criteria to reduce later resource needs.
  \item Most effective when paired with targeted \textit{a posteriori} review, but resource and cost intensive.
\end{tightitemize}
&
\vspace{-0.5\baselineskip}DG1, DG2
\\

\vspace{-0.5\baselineskip}\textbf{Lay Users} &
\begin{tightitemize}
  \item \textit{a posteriori}
\end{tightitemize}
&
\begin{tightitemize}
  \item Develop evaluation criteria in parallel with usability testing while reviewing outputs.
\end{tightitemize}
&
\begin{tightitemize}
  \item Emphasize clarity, readability, format, and tone.
  \item Develop criteria to note gaps or presentation issues.
\end{tightitemize}
&
\begin{tightitemize}
  \item Less emphasis on technical correctness.
  \item Efficient to gather during planned usability testing; low added cost.
\end{tightitemize}
&
\vspace{-0.5\baselineskip}DG3
\\

\vspace{-0.5\baselineskip}\textbf{LLMs} &
\begin{tightitemize}
  \item \textit{a priori}
\end{tightitemize}
&
\begin{tightitemize}
  \item Generate generic criteria ensuring all task requirements in the prompt are addressed.
\end{tightitemize}
&
\begin{tightitemize}
  \item Verify prompt coverage.
  \item Apply established criteria at scale; suggest baseline, generic checks.
  \item Support human evaluators in transforming statements into more objective criteria.
\end{tightitemize}
&
\begin{tightitemize}
  \item Able to manage lower-level evaluation tasks.
  \item If used \textit{a posteriori}, risk of overfitting to or anchoring to flawed outputs.
  \item Risk of bias/hallucination.
  \item Sensitive to prompting strategy and underlying training data.
\end{tightitemize}
&
\vspace{-0.5\baselineskip}DG4, DG5
\\

\bottomrule
\end{tabular}
\end{table*}

\subsection{Do We Experience Criteria Drift or Overfitting to the Data?}

Our findings align with other literature regarding the concept of criteria drift as seen in \citet{shankar2024validates} in that we also observed refinement and addition of new criteria following evaluation of the output. 
Interestingly, we observed that domain experts articulated detailed knowledge-based criteria during the \textit{a priori} phase, displaying their priority in applying knowledge-based criteria at the onset upon reading the prompt.
Following observation of the output, they were more inclined to introduce criteria related to formatting and readability. \revision{Among domain experts and lay users, some sub-themes decreased as new criteria were generated during the output review, which could be because the concrete outputs narrowed participants’ focus and made certain criteria feel more relevant than others.} 
These findings suggest that the outputs had some influence on the criteria development, as participants adapted and expanded their criteria based on their reactions to the content. 
As has been shown in previous work on criteria development, crafting criteria takes multiple iterations as users evaluate outputs \cite{pan2024human, kim2024evallm}. 
Outputs could potentially alter domain experts' mental models of the criteria needed by liking or disliking what was observed after having already established specific criteria based on their knowledge. 
\revision{While lay users exhibited similar behavior, it is less concerning in their case, as their role may rely more on the output to shape criteria that meet their needs in the \textit{a posteriori} phase.}
\delete{While lay users exhibited similar behavior, it is less concerning in their case, as they may rely more on the outputs to set accurate evaluation criteria that meets their needs in the \textit{a posteriori} phase.}

\revision{These patterns can be interpreted through the lens of \textit{belief updating} such that when evaluators encounter new information, they naturally revise their internal standards in response to that evidence~\cite{hogarth1992order}. This process can shift evaluators’ interpretations of what counts as important even when they had already articulated domain-grounded criteria in the \textit{a priori} phase. In addition, belief updating is often directionally shaped by the properties of the shared information source. As Koçak’s sequential updating model shows, when multiple individuals process the same signal, their beliefs tend to shift in similar directions because each update is anchored to the same salient cues in the evidence~\cite{koccak2018sequential}, which may describe the convergence we are seeing in criteria across sources. 
}

Moreover, this issue is not exclusive to human evaluators. The tendency of LLMs to create criteria based on the output during the \textit{a posteriori} phase mirrors the concept of ``overfitting'' found in machine learning, where models become too tailored to their training data and fail to generalize effectively \cite{eigner2024determinants}. This raises significant questions regarding the authenticity of the criteria drift observed in both human participants and LLMs. An important consideration is: \textit{to what extent are we seeing genuine criteria drift, and to what extent are participants and the LLM responding to the specific characteristics of the outputs?} This type of concern should be taken into account when developing tools used for the generation of criteria and the evaluation of the results.

\subsection{Design Implications \revision{and Guidelines} For Evaluation Workflows (RQ3)}

In this section, we draw on the insights learned from our study and address our third research question that discusses the implications for evaluation workflows.
A key implication of our findings is that the evaluation process should be \revision{structured as a staged workflow}\delete{broken down into stages}, in which domain experts, lay users, and LLMs \revision{contribute}\delete{are engaged} at different points to maximize their complementary strengths. Table~\ref{tab:staged_evaluation} summarizes this staged workflow, outlining the recommended roles, timing, and trade-offs for each evaluator type. \revision{In the design guidelines that follow, we describe how each stage builds on prior work in LLM evaluation systems and how a structured workflow can strengthen criteria creation and improve evaluation practices currently employed in industry.}

\paragraph{\revision{Design Guideline 1: Involve Domain Experts in the a Priori Phase.}}\label{DG1}

Our findings show that there is a need to restructure evaluation workflows so that domain experts are centered at the \textit{a priori} stage to set detailed, knowledge-based evaluation criteria that are rooted in professional standards. \revision{In this stage, evaluation workflows should first present experts with only the task scenario or prompt and elicit criteria before exposure to or engagement with any model outputs to ensure that the criteria reflect domain expectations rather than properties of a particular response.}
\revision{Many existing evaluation tools, such as EvalLM, ask evaluators to inspect or grade outputs before criteria are finalized~\cite{kim2024evallm}, which can unintentionally shape or narrow the evaluators’ standards.}
\revision{By contrast, expert-defined \textit{a priori} criteria provide a high-fidelity, domain-grounded rubric that can guide downstream human evaluation or serve as ground truth for training and calibrating LLM-as-a-Judge models~\cite{liu2024calibrating}.}
\revision{This practice aligns with long-standing recommendations in education, where rubrics are created before reviewing student work to reduce unintentional influences on judgment~\cite{malouff2014preventing}. It also mirrors requirements-gathering practices in software engineering, where stakeholders define success criteria independently of system performance~\cite{jarke2010requirements}.}
\revision{Establishing expert criteria upfront therefore strengthens the foundation of the evaluation process, preserves expert time and resources, and supports more consistent, domain-aligned assessment throughout the workflow.}
\delete{Their expertise ensures that complex, domain-specific considerations are addressed early to reduce the risk of bias in later stages and save the time and cost of domain expertise in output evaluation.}

\paragraph{\revision{Design Guideline 2: Support Domain Experts in Reflective Judgment of Outputs in the \textit{a Posteriori} phase}} \label{DG2}

In the \textit{a posteriori} phase, if resources allow, domain experts should return to review outputs for missing content, inaccuracies, or contextual misalignments that only become apparent after outputs are generated. 
\revision{To make the best use of this expertise, the goal of this stage should shift from quick, holistic scoring toward more diagnostic and reflective analysis, which helps ensure that experts’ evaluations are guided by domain standards rather than by features of any single output.}
\revision{Drawing on Norman’s distinction between experiential and reflective cognition~\cite{norman2024things}, this means designing evaluation tasks that encourage slow, deliberate inspection, such as highlighting errors, annotating missing information, or commenting on reasoning flaws, rather than rapid judgments. Experiential judgments could make experts more susceptible to cognitive patterns such as belief updating, in which exposure to the output too early could result in narrow thinking or lower-level considerations. Reflective judgments could instead anchor experts in their original professional expectations.}
\revision{Many existing evaluation systems may unintentionally push evaluators into experiential cognition by asking them to quickly grade, rank, or choose between outputs~\cite{Wiltberger_2025, chiang2024chatbot, kim2024evallm}. By contrast, workflows that prompt experts to engage in reflective analysis could surface richer insights about correctness, context, and domain alignment. }

\paragraph{\revision{Design Guideline 3: Incorporate Lay Users’ Evaluation a Posteriori Concurrently with Usability Testing}}

Lay users, who focus their evaluation criteria on \delete{usability, such as }formatting, readability, and presentation, may not require separate large-scale data annotation efforts. Our findings suggest that their priorities can be gathered during usability testing in the \textit{a posteriori} phase. \revision{Prior work in usability testing shows that end users are best positioned to identify issues related to clarity, workflow, comprehensibility, and interface expectations, rather than domain-specific correctness~\cite{huovinen2024assessing}.} Since usability testing is a common strategy in LLM tool development, and lay users will likely be evaluating outputs in this process, this offers an efficient way to capture and incorporate their priorities without significant additional cost. \revision{While lay users may also experience cognitive patterns such as belief updating when assessing outputs, this is less concerning for this group, as their role is to react to and refine criteria based on the output experience rather than maintain fixed domain standards.} The criteria they generate can be directly integrated into the overall evaluation framework to ensure that usability considerations are assessed alongside technical accuracy.

\paragraph{\revision{Design Guideline 4: Use LLMs for Lower-Level Tasks a Priori}}

\delete{Given the potential for overfitting, LLMs should not be used to generate criteria in the \textit{a posteriori} phase.}
\revision{Based on our results}, there is an opportunity to use LLMs \delete{in tandem with domain experts}to generate \revision{criteria }during the \textit{a priori} phase \revision{for lower-level tasks} such as ensuring the output meets the prompt's specific high-level task requirements. \revision{When used in this stage, prompts should be designed so that the LLM does not receive or infer information from specific outputs to help avoid anchoring effects.} \revision{This would be particularly useful} if involving domain experts is not feasible.\delete{, we recommend using LLMs to manage lower-level tasks} 
By assigning routine tasks to LLMs and reserving more complex evaluations for domain experts, the evaluation process could become more cost-effective and balanced. 
\revision{In addition, if criteria developed by domain experts or lay-user usability insights is available, it can be incorporated into the prompt to help the LLM generate stronger \textit{a priori} criteria aligned with human standards.}
By integrating these criteria into the training data or prompt, designers can work toward aligning LLMs more closely with the standards and expectations of both domain experts and end-users. 
For example, the LLMCRIT framework demonstrates how using criteria derived from expert guidelines can help models generate more structured and actionable feedback, making it a promising approach for fine-tuning LLM tasks toward expert-level expectations \cite{lubos2024leveraging}. 

\paragraph{\revision{Design Guideline 5: Use LLMs a Priori to Support Objective Criteria}}

\revision{Our findings indicate that evaluation workflows should include LLM systems to support the creation of criteria that are well-defined and support reliable and reproducible assessment~\cite{naik2011software}.} 
We observed that although participants often expressed preferences about the outputs, they found it difficult to translate these reactions into clear assertion statements, \revision{which supports prior work showing that users frequently lack mental models for generating objective evaluative criteria~\cite{zamfirescu2023johnny, liu2022design}.}
\revision{When these criteria that are overly broad or loosely specified are ultimately used, they may result in reduced reliability for LLM-as-a-Judge scoring~\cite{peng2024survey}.} 
\delete{This process sometimes limited the creation of definitive evaluation criteria that could be expressed as clear assertion statements.}

\revision{In the early stages, where \textit{a priori} criteria are drafted, LLMs can help users move from vague criteria to clearer and more actionable evaluation standards.}\delete{This suggests should support the development of criteria, perhaps by offering guidance or tools that help users articulate their evaluation standards more clearly.} \revision{Tools such as MetricMate, which provides reflection prompts~\cite{gebreegziabher2025metricmate}, Evalet, which supports functional fragmentation~\cite{kim2025evalet}, and BloomIntent, which offers intent generation and clustering~\cite{choi2025bloomintent}, have shown how interactive features using LLMs can help users develop clearer and more structured evaluation criteria. Evaluation workflows can draw on these approaches to help produce more objective criteria for downstream LLM-as-a-Judge pipelines.}
\delete{Tools, such as MetricMate, have developed interactive features to encourage users to reflect and iterate on the criteria to achieve more specific assertions~\cite{gebreegziabher2025metricmate}.} \delete{Using these tools would streamline the evaluation process and ensure that both groups contribute to the refinement of LLM outputs.} 

\revision{However, our findings also suggest that not all criteria may need to be maximally objective. Participants occasionally left the criteria broad, and in these cases, ambiguity was not a defect but a deliberate choice that signaled where further human discussion was essential. This reframes the role of criteria in human-in-the-loop evaluation workflows such that, rather than functioning solely as grading instruments for LLM-as-a-Judge systems, criteria can also guide evaluators, structure discussion, and highlight ambiguous interpretations. Intentionally high-level criteria leave room for judgment and pinpoint the complexities of the tasks which may otherwise be lost when all ambiguities are forced into assertions. Future workflows should therefore support using the LLM to enable reflection and deliberation among human evaluators.}

\section{Limitations and Future Work}
Our work has some limitations that can lead to exciting future explorations. 
\revision{First, our study involves a relatively small number of domain experts and lay users. This sample size is consistent with prior qualitative work examining expert reasoning or evaluation practices~\cite{szymanski2024limitations}, yet it necessarily limits broad generalizability. Because our aim was to generate process-level insights into how criteria emerge and evolve, our approach aligns with interpretive research traditions that prioritize contextual depth and situated understanding~\cite{soden2024evaluating}. Within this framing, the limited sample allowed us to closely examine the mechanisms through which different sources construct and iteratively refine evaluation criteria. However, we acknowledge that criteria-formation processes may vary in domains with different expertise or tasks, and future work should examine additional domains and larger participant pools to understand how these patterns scale.} 
\delete{First, this study examined two domains of nutrition and math pedagogy to illustrate how different groups contribute to the evaluation process. While our findings demonstrate the framework’s applicability in these domains, future work could explore how evaluation workflows might be tailored to the specific needs, technical requirements, or norms of other domains, potentially leading to domain-specific adaptations that enhance effectiveness. }

Second, while our findings suggest \revision{design guidelines for} a new evaluation workflow, it has yet to be tested in practical settings. Future work could focus on building systems and tools that implement and support this workflow to ensure that it is practical and scalable for real-world applications. Implementing our proposed evaluation strategies in live systems will allow further refinement based on practical challenges and insights gathered during usage.  \revision{In addition, while our findings offered interesting insights into the differences in evaluation criteria, we did not test if the evaluation criteria created by different sources lead to more reliable scores compared to prior rubrics. Future work should evaluate how different sources are interpreted with the LLM-as-a-Judge systems and if LLM judgments would align with human judgments.}

\revision{Third, our study adopts a scenario-specific evaluation approach centered on a set of complex, domain-specific prompt instructions that reflect the knowledge-intensive nature of many LLM applications. While this differs from evaluating a collection of related prompts as seen in other evaluation systems, our findings provide the empirical foundation necessary to understand criteria formation and evolution in domain-specific contexts. Future work could explore how criteria generalize across multiple related prompts and systematically test the design implications of applying the staged approach to broader task families.}

Additionally, we recognize the limitations surrounding the replicability of our findings due to potential changes in LLMs over time. The use of LLMs in our study to generate criteria and outputs introduces the challenge of evolving models, which may affect the consistency of the results if the study is replicated at a later date. The results and evaluation criteria used in our study were generated in August 2024, and updates or changes in the LLMs may lead to variations in performance or outcomes. 

\delete{While this study focused on the evaluation process, there may be other opportunities to investigate how criteria-setting can be further optimized by integrating domain experts, lay users, and LLMs into other stages of model development, such as working to use these criteria to improve the model during prompt generation or model fine-tuning.  All of these approaches could incorporate domain expertise to improve the model.}

Our findings also highlight the potential to utilize domain expert-established criteria as benchmarks not only to evaluate, but also to refine LLM outputs. Previous literature has used comprehensive criteria to generate natural language feedback for task execution \cite{lubos2024leveraging}. Criteria developed by domain experts or lay users can serve as foundational elements in the LLM training and fine-tuning processes to improve output alignment in future work.

Finally, while our study demonstrates the complementary strengths of domain experts, lay users, and LLMs, further research is needed to extend across other domains that require expertise and are complex in nature. Our findings suggest that different sources of input may be most effective at specific stages of the evaluation workflow. Future work should focus on developing comprehensive frameworks to guide AI developers in systematically designing evaluation processes. These frameworks could further identify the most effective stages for incorporating input from each stakeholder and determine the key contributions each source can provide to ensure robust and reliable outcomes.

\vspace{-1.8pt}

\section{Conclusion}
This paper examined how domain experts, lay users, and LLMs contribute to the development of evaluation criteria for complex, domain-specific tasks, with particular attention to how these criteria evolve across the \textit{a priori} and \textit{a posteriori} phases of evaluation. Our findings reveal distinct patterns in the motivations and contributions of each source, and reveal complementary strengths that can be applied within the evaluation process. In addition, our results confirm the presence of criteria drift and show convergence in the \textit{a posteriori} phase, where all sources tend to focus on observable features in the output. While this shared focus may support consistency, it also shows the importance of involving human judgment earlier in the process to ensure long-term quality and reduce the risk of reinforcing flawed model behavior. Based on these insights, we propose \revision{design guidelines for} a staged evaluation workflow that integrates each source according to its strengths to balance cost and scalability.

\begin{acks}
This work was supported by the Agriculture and Food Research Initiative grant no. 2021-67022-33447/project accession no. 1024822 from the USDA National Institute of Food and Agriculture.
\end{acks}



\bibliographystyle{ACM-Reference-Format}
\bibliography{base.bib}

@inproceedings{szymanski2024integrating,
  title={Integrating Expertise in LLMs: Crafting a Customized Nutrition Assistant with Refined Template Instructions},
  author={Szymanski, Annalisa and Wimer, Brianna L and Anuyah, Oghenemaro and Eicher-Miller, Heather A and Metoyer, Ronald A},
  booktitle={Proceedings of the CHI Conference on Human Factors in Computing Systems},
  pages={1--22},
  year={2024}
}

@inproceedings{ma2025towards,
author = {Ma, Shuai and Chen, Qiaoyi and Wang, Xinru and Zheng, Chengbo and Peng, Zhenhui and Yin, Ming and Ma, Xiaojuan},
title = {Towards Human-AI Deliberation: Design and Evaluation of LLM-Empowered Deliberative AI for AI-Assisted Decision-Making},
year = {2025},
isbn = {9798400713941},
publisher = {Association for Computing Machinery},
address = {New York, NY, USA},
url = {https://doi.org/10.1145/3706598.3713423},
doi = {10.1145/3706598.3713423},
booktitle = {Proceedings of the 2025 CHI Conference on Human Factors in Computing Systems},
articleno = {261},
numpages = {23},
location = {
},
series = {CHI '25}
}

@article{spitzer2025human,
  title={Human Delegation Behavior in Human-AI Collaboration: The Effect of Contextual Information},
  author={Spitzer, Philipp and Holstein, Joshua and Hemmer, Patrick and V{\"o}ssing, Michael and K{\"u}hl, Niklas and Martin, Dominik and Satzger, Gerhard},
  journal={Proceedings of the ACM on Human-Computer Interaction},
  volume={9},
  number={2},
  pages={1--28},
  year={2025},
  publisher={ACM New York, NY, USA}
}

@article{schafer2025ai,
  title={'The AI is uncertain, so am I. What now?': Navigating Shortcomings of Uncertainty Representations in Human-AI Collaboration with Capability-focused Guidance},
  author={Sch{\"a}fer, Ulrike and Sipos, Lars and M{\"u}ller-Birn, Claudia},
  journal={Proceedings of the ACM on Human-Computer Interaction},
  volume={9},
  number={7},
  pages={1--48},
  year={2025},
  publisher={ACM New York, NY, USA}
}

@inproceedings{umbelino2025an,
author = {Kreia Umbelino, Gustavo and Veri, Francesco},
title = {An Emergent Understanding of Human-AI Collaboration in Deliberation},
year = {2025},
isbn = {9798400714801},
publisher = {Association for Computing Machinery},
address = {New York, NY, USA},
url = {https://doi.org/10.1145/3715070.3749265},
doi = {10.1145/3715070.3749265},
booktitle = {Companion Publication of the 2025 Conference on Computer-Supported Cooperative Work and Social Computing},
pages = {426–434},
numpages = {9},
location = {
},
series = {CSCW Companion '25}
}

@inproceedings{wang2025aideation,
author = {Wang, Wen-Fan and Lu, Chien-Ting and Ponsa i Campany\`{a}, Nil and Chen, Bing-Yu and Chen, Mike Y.},
title = {AIdeation: Designing a Human-AI Collaborative Ideation System for Concept Designers},
year = {2025},
isbn = {9798400713941},
publisher = {Association for Computing Machinery},
address = {New York, NY, USA},
url = {https://doi.org/10.1145/3706598.3714148},
doi = {10.1145/3706598.3714148},
booktitle = {Proceedings of the 2025 CHI Conference on Human Factors in Computing Systems},
articleno = {21},
numpages = {28},
location = {
},
series = {CHI '25}
}

@article{soden2024evaluating,
  author = {Soden, Robert and Toombs, Austin and Thomas, Michaelanne},
title = {Evaluating Interpretive Research in HCI},
year = {2024},
issue_date = {January - February 2024},
publisher = {Association for Computing Machinery},
address = {New York, NY, USA},
volume = {31},
number = {1},
issn = {1072-5520},
url = {https://doi.org/10.1145/3633200},
doi = {10.1145/3633200},
journal = {Interactions},
month = jan,
pages = {38–42},
numpages = {5}
}

@article{gao2024llm,
  title={Llm-based nlg evaluation: Current status and challenges},
  author={Gao, Mingqi and Hu, Xinyu and Yin, Xunjian and Ruan, Jie and Pu, Xiao and Wan, Xiaojun},
  journal={Computational Linguistics},
  pages={1--27},
  year={2025},
  publisher={MIT Press 255 Main Street, 9th Floor, Cambridge, Massachusetts 02142, USA~…}
}

@inproceedings{shankar2024validates,
  title={Who validates the validators? aligning llm-assisted evaluation of llm outputs with human preferences},
  author={Shankar, Shreya and Zamfirescu-Pereira, JD and Hartmann, Bj{\"o}rn and Parameswaran, Aditya and Arawjo, Ian},
  booktitle={Proceedings of the 37th Annual ACM Symposium on User Interface Software and Technology},
  pages={1--14},
  year={2024}
}

@article{zheng2024judging,
  title={Judging llm-as-a-judge with mt-bench and chatbot arena},
  author={Zheng, Lianmin and Chiang, Wei-Lin and Sheng, Ying and Zhuang, Siyuan and Wu, Zhanghao and Zhuang, Yonghao and Lin, Zi and Li, Zhuohan and Li, Dacheng and Xing, Eric and others},
  journal={Advances in Neural Information Processing Systems},
  volume={36},
  year={2024}
}

@article{khashabi2021genie,
  title={GENIE: Toward reproducible and standardized human evaluation for text generation},
  author={Khashabi, Daniel and Stanovsky, Gabriel and Bragg, Jonathan and Lourie, Nicholas and Kasai, Jungo and Choi, Yejin and Smith, Noah A and Weld, Daniel S},
  journal={arXiv preprint arXiv:2101.06561},
  year={2021}
}

@inproceedings{papineni2002bleu,
  title={Bleu: a method for automatic evaluation of machine translation},
  author={Papineni, Kishore and Roukos, Salim and Ward, Todd and Zhu, Wei-Jing},
  booktitle={Proceedings of the 40th annual meeting of the Association for Computational Linguistics},
  pages={311--318},
  year={2002}
}

@inproceedings{lin2004rouge,
  title={Rouge: A package for automatic evaluation of summaries},
  author={Lin, Chin-Yew},
  booktitle={Text summarization branches out},
  pages={74--81},
  year={2004}
}

@article{gehrmann2023repairing,
  title={Repairing the cracked foundation: A survey of obstacles in evaluation practices for generated text},
  author={Gehrmann, Sebastian and Clark, Elizabeth and Sellam, Thibault},
  journal={Journal of Artificial Intelligence Research},
  volume={77},
  pages={103--166},
  year={2023}
}

@inproceedings{arawjo2024chainforge,
  title={ChainForge: A Visual Toolkit for Prompt Engineering and LLM Hypothesis Testing},
  author={Arawjo, Ian and Swoopes, Chelse and Vaithilingam, Priyan and Wattenberg, Martin and Glassman, Elena L},
  booktitle={Proceedings of the CHI Conference on Human Factors in Computing Systems},
  pages={1--18},
  year={2024}
}

@inproceedings{lieb2024student,
  title={Student Interaction with NewtBot: An LLM-as-tutor Chatbot for Secondary Physics Education},
  author={Lieb, Anna and Goel, Toshali},
  booktitle={Extended Abstracts of the CHI Conference on Human Factors in Computing Systems},
  pages={1--8},
  year={2024}
}

@inproceedings{kim2024evallm,
  title={Evallm: Interactive evaluation of large language model prompts on user-defined criteria},
  author={Kim, Tae Soo and Lee, Yoonjoo and Shin, Jamin and Kim, Young-Ho and Kim, Juho},
  booktitle={Proceedings of the CHI Conference on Human Factors in Computing Systems},
  pages={1--21},
  year={2024}
}

@article{chen2024designing,
  title={Designing a Dashboard for Transparency and Control of Conversational AI},
  author={Chen, Yida and Wu, Aoyu and DePodesta, Trevor and Yeh, Catherine and Li, Kenneth and Marin, Nicholas Castillo and Patel, Oam and Riecke, Jan and Raval, Shivam and Seow, Olivia and others},
  journal={arXiv preprint arXiv:2406.07882},
  year={2024}
}

@article{marjaei2019maxqda,
  title={MAXQDA and its Application to LIS Research},
  author={Marjaei, Seyedhadi and Yazdi, Fahimeh Ahmadian and Chandrashekara, M},
  journal={Library Philosophy and Practice},
  pages={1--9},
  year={2019},
  publisher={Library Philosophy and Practice}
}

@article{sarkar2024llms,
  title={LLMs as On-demand Customizable Service},
  author={Sarkar, Souvika and Babar, Mohammad Fakhruddin and Hasan, Monowar and Karmaker, Shubhra Kanti},
  journal={arXiv preprint arXiv:2401.16577},
  year={2024}
}

@inproceedings{sallam2023chatgpt,
  title={ChatGPT utility in healthcare education, research, and practice: systematic review on the promising perspectives and valid concerns},
  author={Sallam, Malik},
  booktitle={Healthcare},
  volume={11},
  number={6},
  pages={887},
  year={2023},
  organization={MDPI}
}

@article{ahmad2023creating,
  title={Creating trustworthy llms: Dealing with hallucinations in healthcare ai},
  author={Ahmad, Muhammad Aurangzeb and Yaramis, Ilker and Roy, Taposh Dutta},
  journal={arXiv preprint arXiv:2311.01463},
  year={2023}
}

@article{chatelan2023chatgpt,
  title={ChatGPT and future artificial intelligence chatbots: what may be the influence on credentialed nutrition and dietetics practitioners?},
  author={Chatelan, Angeline and Clerc, Aur{\'e}lien and Fonta, Pierre-Alexandre},
  journal={Journal of the Academy of Nutrition and Dietetics},
  volume={123},
  number={11},
  pages={1525--1531},
  year={2023}
}

@article{ayers2023comparing,
  title={Comparing physician and artificial intelligence chatbot responses to patient questions posted to a public social media forum},
  author={Ayers, John W and Poliak, Adam and Dredze, Mark and Leas, Eric C and Zhu, Zechariah and Kelley, Jessica B and Faix, Dennis J and Goodman, Aaron M and Longhurst, Christopher A and Hogarth, Michael and others},
  journal={JAMA internal medicine},
  volume={183},
  number={6},
  pages={589--596},
  year={2023},
  publisher={American Medical Association}
}

@article{wang2024prompt,
  title={Prompt engineering in consistency and reliability with the evidence-based guideline for LLMs},
  author={Wang, Li and Chen, Xi and Deng, XiangWen and Wen, Hao and You, MingKe and Liu, WeiZhi and Li, Qi and Li, Jian},
  journal={npj Digital Medicine},
  volume={7},
  number={1},
  pages={41},
  year={2024},
  publisher={Nature Publishing Group UK London}
}

@article{andriopoulos2023augmenting,
  title={Augmenting LLMs with Knowledge: A survey on hallucination prevention},
  author={Andriopoulos, Konstantinos and Pouwelse, Johan},
  journal={arXiv preprint arXiv:2309.16459},
  year={2023}
}

@article{liu2024exploring,
  title={Exploring and evaluating hallucinations in llm-powered code generation},
  author={Liu, Fang and Liu, Yang and Shi, Lin and Huang, Houkun and Wang, Ruifeng and Yang, Zhen and Zhang, Li},
  journal={arXiv preprint arXiv:2404.00971},
  year={2024}
}

@article{pan2024human,
  title={Human-Centered Design Recommendations for LLM-as-a-Judge},
  author={Pan, Qian and Ashktorab, Zahra and Desmond, Michael and Cooper, Martin Santillan and Johnson, James and Nair, Rahul and Daly, Elizabeth and Geyer, Werner},
  journal={arXiv preprint arXiv:2407.03479},
  year={2024}
}

@article{wang2024understanding,
  title={Understanding User Experience in Large Language Model Interactions},
  author={Wang, Jiayin and Ma, Weizhi and Sun, Peijie and Zhang, Min and Nie, Jian-Yun},
  journal={arXiv preprint arXiv:2401.08329},
  year={2024}
}

@article{jiao2024navigating,
  title={Navigating llm ethics: Advancements, challenges, and future directions},
  author={Jiao, Junfeng and Afroogh, Saleh and Xu, Yiming and Phillips, Connor},
  journal={arXiv preprint arXiv:2406.18841},
  year={2024}
}

@article{gururangan2020don,
  title={Don't stop pretraining: Adapt language models to domains and tasks},
  author={Gururangan, Suchin and Marasovi{\'c}, Ana and Swayamdipta, Swabha and Lo, Kyle and Beltagy, Iz and Downey, Doug and Smith, Noah A},
  journal={arXiv preprint arXiv:2004.10964},
  year={2020}
}

@inproceedings{cheong2024not,
  title={(A) I Am Not a Lawyer, But...: Engaging Legal Experts towards Responsible LLM Policies for Legal Advice},
  author={Cheong, Inyoung and Xia, King and Feng, KJ Kevin and Chen, Quan Ze and Zhang, Amy X},
  booktitle={The 2024 ACM Conference on Fairness, Accountability, and Transparency},
  pages={2454--2469},
  year={2024}
}

@article{ziegler2019fine,
  title={Fine-tuning language models from human preferences},
  author={Ziegler, Daniel M and Stiennon, Nisan and Wu, Jeffrey and Brown, Tom B and Radford, Alec and Amodei, Dario and Christiano, Paul and Irving, Geoffrey},
  journal={arXiv preprint arXiv:1909.08593},
  year={2019}
}

@article{kirk2023comparison,
  title={Comparison of answers between ChatGPT and human dieticians to common nutrition questions},
  author={Kirk, Daniel and van Eijnatten, Elise and Camps, Guido},
  journal={Journal of Nutrition and Metabolism},
  volume={2023},
  number={1},
  pages={5548684},
  year={2023},
  publisher={Wiley Online Library}
}

@article{ponzo2024chatgpt,
  title={Is ChatGPT an Effective Tool for Providing Dietary Advice?},
  author={Ponzo, Valentina and Goitre, Ilaria and Favaro, Enrica and Merlo, Fabio Dario and Mancino, Maria Vittoria and Riso, Sergio and Bo, Simona},
  journal={Nutrients},
  volume={16},
  number={4},
  pages={469},
  year={2024},
  publisher={MDPI}
}

@article{garcia2023chatgpt,
  title={ChatGPT as a virtual dietitian: Exploring its potential as a tool for improving nutrition knowledge},
  author={Garcia, Manuel B},
  journal={Applied System Innovation},
  volume={6},
  number={5},
  pages={96},
  year={2023},
  publisher={MDPI}
}

@inproceedings{zhang2024llmeval,
  title={Llmeval: A preliminary study on how to evaluate large language models},
  author={Zhang, Yue and Zhang, Ming and Yuan, Haipeng and Liu, Shichun and Shi, Yongyao and Gui, Tao and Zhang, Qi and Huang, Xuanjing},
  booktitle={Proceedings of the AAAI Conference on Artificial Intelligence},
  volume={38},
  number={17},
  pages={19615--19622},
  year={2024}
}

@misc{webster2023promptfoo,
  title={promptfoo: Test your prompts},
  author={Webster, Ian},
  year={2023},
  howpublished={\url{https://www.promptfoo.dev/}},
}

@article{eigner2024determinants,
  title={Determinants of llm-assisted decision-making},
  author={Eigner, Eva and H{\"a}ndler, Thorsten},
  journal={arXiv preprint arXiv:2402.17385},
  year={2024}
}

@article{wang2024large,
  title={Large language models cannot replace human participants because they cannot portray identity groups},
  author={Wang, Angelina and Morgenstern, Jamie and Dickerson, John P},
  journal={arXiv preprint arXiv:2402.01908},
  year={2024}
}

@inproceedings{hamalainen2023evaluating,
  title={Evaluating large language models in generating synthetic hci research data: a case study},
  author={H{\"a}m{\"a}l{\"a}inen, Perttu and Tavast, Mikke and Kunnari, Anton},
  booktitle={Proceedings of the 2023 CHI Conference on Human Factors in Computing Systems},
  pages={1--19},
  year={2023}
}

@article{gilardi2023chatgpt,
  title={ChatGPT outperforms crowd workers for text-annotation tasks},
  author={Gilardi, Fabrizio and Alizadeh, Meysam and Kubli, Ma{\"e}l},
  journal={Proceedings of the National Academy of Sciences},
  volume={120},
  number={30},
  pages={e2305016120},
  year={2023},
  publisher={National Acad Sciences}
}

@article{dong2024can,
  title={Can LLM be a Personalized Judge?},
  author={Dong, Yijiang River and Hu, Tiancheng and Collier, Nigel},
  journal={arXiv preprint arXiv:2406.11657},
  year={2024}
}

@inproceedings{lubos2024leveraging,
  title={Leveraging LLMs for the Quality Assurance of Software Requirements},
  author={Lubos, Sebastian and Felfernig, Alexander and Tran, Thi Ngoc Trang and Garber, Damian and El Mansi, Merfat and Erdeniz, Seda Polat and Le, Viet-Man},
  booktitle={2024 IEEE 32nd International Requirements Engineering Conference (RE)},
  pages={389--397},
  year={2024},
  organization={IEEE}
}

@inproceedings{kalai2024calibrated,
  title={Calibrated language models must hallucinate},
  author={Kalai, Adam Tauman and Vempala, Santosh S},
  booktitle={Proceedings of the 56th Annual ACM Symposium on Theory of Computing},
  pages={160--171},
  year={2024}
}

@article{liu2023trustworthy,
  title={Trustworthy LLMs: A survey and guideline for evaluating large language models' alignment},
  author={Liu, Yang and Yao, Yuanshun and Ton, Jean-Francois and Zhang, Xiaoying and Cheng, Ruocheng Guo Hao and Klochkov, Yegor and Taufiq, Muhammad Faaiz and Li, Hang},
  journal={arXiv preprint arXiv:2308.05374},
  year={2023}
}

@article{huang2024trustllm,
  title={Trustllm: Trustworthiness in large language models},
  author={Huang, Yue and Sun, Lichao and Wang, Haoran and Wu, Siyuan and Zhang, Qihui and Li, Yuan and Gao, Chujie and Huang, Yixin and Lyu, Wenhan and Zhang, Yixuan and others},
  journal={arXiv preprint arXiv:2401.05561},
  year={2024}
}

@article{liao2023ai,
  title={Ai transparency in the age of llms: A human-centered research roadmap},
  author={Liao, Q Vera and Vaughan, Jennifer Wortman},
  journal={arXiv preprint arXiv:2306.01941},
  pages={5368--5393},
  year={2023},
  publisher={no}
}

@article{liusie2024efficient,
  title={Efficient LLM Comparative Assessment: a Product of Experts Framework for Pairwise Comparisons},
  author={Liusie, Adian and Raina, Vatsal and Fathullah, Yassir and Gales, Mark},
  journal={arXiv preprint arXiv:2405.05894},
  year={2024}
}

@article{sun2024lalaeval,
  title={LalaEval: A Holistic Human Evaluation Framework for Domain-Specific Large Language Models},
  author={Sun, Chongyan and Lin, Ken and Wang, Shiwei and Wu, Hulong and Fu, Chengfei and Wang, Zhen},
  journal={arXiv preprint arXiv:2408.13338},
  year={2024}
}

@article{williams2019art,
  title={The art of coding and thematic exploration in qualitative research},
  author={Williams, Michael and Moser, Tami},
  journal={International management review},
  volume={15},
  number={1},
  pages={45--55},
  year={2019}
}

@article{cheng2023now,
  title={The now and future of ChatGPT and GPT in psychiatry},
  author={Cheng, Szu-Wei and Chang, Chung-Wen and Chang, Wan-Jung and Wang, Hao-Wei and Liang, Chih-Sung and Kishimoto, Taishiro and Chang, Jane Pei-Chen and Kuo, John S and Su, Kuan-Pin},
  journal={Psychiatry and clinical neurosciences},
  volume={77},
  number={11},
  pages={592--596},
  year={2023},
  publisher={Wiley Online Library}
}

@article{king2023introduction,
  title={An introduction to generative artificial intelligence in mental health care: considerations and guidance},
  author={King, Darlene R and Nanda, Guransh and Stoddard, Joel and Dempsey, Allison and Hergert, Sarah and Shore, Jay H and Torous, John},
  journal={Current psychiatry reports},
  volume={25},
  number={12},
  pages={839--846},
  year={2023},
  publisher={Springer}
}

@article{liu2023chatcounselor,
  title={Chatcounselor: A large language models for mental health support},
  author={Liu, June M and Li, Donghao and Cao, He and Ren, Tianhe and Liao, Zeyi and Wu, Jiamin},
  journal={arXiv preprint arXiv:2309.15461},
  year={2023}
}

@article{wang2023assessing,
  title={Assessing the reliability of large language model knowledge},
  author={Wang, Weixuan and Haddow, Barry and Birch, Alexandra and Peng, Wei},
  journal={arXiv preprint arXiv:2310.09820},
  year={2023}
}

@article{ji2023survey,
  title={Survey of hallucination in natural language generation},
  author={Ji, Ziwei and Lee, Nayeon and Frieske, Rita and Yu, Tiezheng and Su, Dan and Xu, Yan and Ishii, Etsuko and Bang, Ye Jin and Madotto, Andrea and Fung, Pascale},
  journal={ACM Computing Surveys},
  volume={55},
  number={12},
  pages={1--38},
  year={2023},
  publisher={ACM New York, NY}
}

@article{huang2023survey,
  title={A survey on hallucination in large language models: Principles, taxonomy, challenges, and open questions},
  author={Huang, Lei and Yu, Weijiang and Ma, Weitao and Zhong, Weihong and Feng, Zhangyin and Wang, Haotian and Chen, Qianglong and Peng, Weihua and Feng, Xiaocheng and Qin, Bing and others},
  journal={ACM Transactions on Information Systems},
  year={2023},
  publisher={ACM New York, NY}
}

@article{sonkar2024pedagogical,
  title={Pedagogical alignment of large language models},
  author={Sonkar, Shashank and Ni, Kangqi and Chaudhary, Sapana and Baraniuk, Richard G},
  journal={arXiv preprint arXiv:2402.05000},
  year={2024}
}

@article{lu2024ahp,
  title={AHP-Powered LLM Reasoning for Multi-Criteria Evaluation of Open-Ended Responses},
  author={Lu, Xiaotian and Li, Jiyi and Takeuchi, Koh and Kashima, Hisashi},
  journal={arXiv preprint arXiv:2410.01246},
  year={2024}
}

@inproceedings{wang2021seeing,
  title={Seeing beyond expert blind spots: Online learning design for scale and quality},
  author={Wang, Xu and Rose, Carolyn and Koedinger, Ken},
  booktitle={Proceedings of the 2021 CHI Conference on Human Factors in Computing Systems},
  pages={1--14},
  year={2021}
}

@article{wu2024aligning,
  title={Aligning LLMs with Individual Preferences via Interaction},
  author={Wu, Shujin and Fung, May and Qian, Cheng and Kim, Jeonghwan and Hakkani-Tur, Dilek and Ji, Heng},
  journal={arXiv preprint arXiv:2410.03642},
  year={2024}
}

@article{kirk2024benefits,
  title={The benefits, risks and bounds of personalizing the alignment of large language models to individuals},
  author={Kirk, Hannah Rose and Vidgen, Bertie and R{\"o}ttger, Paul and Hale, Scott A},
  journal={Nature Machine Intelligence},
  pages={1--10},
  year={2024},
  publisher={Nature Publishing Group UK London}
}

@article{gabriel2022challenge,
  title={The challenge of value alignment},
  author={Gabriel, Iason and Ghazavi, Vafa},
  journal={The Oxford handbook of digital ethics},
  pages={336--355},
  year={2022},
  publisher={Oxford University Press Oxford}
}

@article{chen2024self,
  title={Self-play fine-tuning converts weak language models to strong language models},
  author={Chen, Zixiang and Deng, Yihe and Yuan, Huizhuo and Ji, Kaixuan and Gu, Quanquan},
  journal={arXiv preprint arXiv:2401.01335},
  year={2024}
}

@inproceedings{han2023robustqa,
  title={RobustQA: Benchmarking the robustness of domain adaptation for open-domain question answering},
  author={Han, Rujun and Qi, Peng and Zhang, Yuhao and Liu, Lan and Burger, Juliette and Wang, William Yang and Huang, Zhiheng and Xiang, Bing and Roth, Dan},
  booktitle={Findings of the Association for Computational Linguistics: ACL 2023},
  pages={4294--4311},
  year={2023}
}

@article{zhu2021retrieving,
  title={Retrieving and reading: A comprehensive survey on open-domain question answering},
  author={Zhu, Fengbin and Lei, Wenqiang and Wang, Chao and Zheng, Jianming and Poria, Soujanya and Chua, Tat-Seng},
  journal={arXiv preprint arXiv:2101.00774},
  year={2021}
}

@inproceedings{arawjo2023chainforge,
  title={ChainForge: An open-source visual programming environment for prompt engineering},
  author={Arawjo, Ian and Vaithilingam, Priyan and Wattenberg, Martin and Glassman, Elena},
  booktitle={Adjunct Proceedings of the 36th Annual ACM Symposium on User Interface Software and Technology},
  pages={1--3},
  year={2023}
}

@article{Tokayev_2023, title={Ethical Implications of Large Language Models A Multidimensional Exploration of Societal, Economic, and Technical Concerns}, volume={8}, url={https://norislab.com/index.php/ijsa/article/view/42}, number={9}, journal={International Journal of Social Analytics}, author={Tokayev, Kassym-Jomart}, year={2023}, month={Sep.}, pages={17–33} }

@inproceedings{zhang2024way,
  title={Way to specialist: Closing loop between specialized llm and evolving domain knowledge graph},
  author={Zhang, Yutong and Chen, Lixing and Li, Shenghong and Cao, Nan and Shi, Yang and Ding, Jiaxin and Qu, Zhe and Zhou, Pan and Bai, Yang},
  booktitle={Proceedings of the 31st ACM SIGKDD Conference on Knowledge Discovery and Data Mining V. 1},
  pages={1996--2007},
  year={2025}
}

@inproceedings{szymanski2024limitations,
  title={Limitations of the llm-as-a-judge approach for evaluating llm outputs in expert knowledge tasks},
  author={Szymanski, Annalisa and Ziems, Noah and Eicher-Miller, Heather A and Li, Toby Jia-Jun and Jiang, Meng and Metoyer, Ronald A},
  booktitle={Proceedings of the 30th International Conference on Intelligent User Interfaces},
  pages={952--966},
  year={2025}
}

@inproceedings{sun2024trust,
  title={Trust by Interface: How Different User Interfaces Shape Human Trust in Health Information from Large Language Models},
  author={Sun, Xin and Liu, Yunjie and De Wit, Jan and Bosch, Jos A and Li, Zhuying},
  booktitle={Extended Abstracts of the CHI Conference on Human Factors in Computing Systems},
  pages={1--7},
  year={2024}
}

@inproceedings{gebreegziabher2024supporting,
  title={Supporting Co-Adaptive Machine Teaching through Human Concept Learning and Cognitive Theories},
  author={Gebreegziabher, Simret Araya and Yang, Yukun and Glassman, Elena L and Li, Toby Jia-Jun},
  booktitle={Proceedings of the 2025 CHI Conference on Human Factors in Computing Systems},
  pages={1--18},
  year={2025}
}

@article{gebreegziabher2024leveraging,
  title={Leveraging Variation Theory in Counterfactual Data Augmentation for Optimized Active Learning},
  author={Gebreegziabher, Simret Araya and Ai, Kuangshi and Zhang, Zheng and Glassman, Elena L and Li, Toby Jia-Jun},
  journal={arXiv preprint arXiv:2408.03819},
  year={2024}
}

@article{roseman2021academy,
  title={Academy of nutrition and dietetics: revised 2021 standards of professional performance for registered dietitian nutritionists (competent, proficient, and expert) in management of food and nutrition systems},
  author={Roseman, Mary G and Miller, Sandra N},
  journal={Journal of the Academy of Nutrition and Dietetics},
  volume={121},
  number={6},
  pages={1157--1174},
  year={2021},
  publisher={Elsevier}
}

@article{wood2015exploring,
  title={Exploring dietitians’ engagement with health literacy: concept and practice},
  author={Wood, Jennifer and Gillis, Doris E},
  journal={Canadian Journal of Dietetic Practice and Research},
  volume={76},
  number={2},
  pages={51--55},
  year={2015},
  publisher={Dietitians of Canada}
}

@misc{miller2024llm,
  title={LLM Based Math Tutoring: Challenges and Dataset},
  author={Miller, Pepper and DiCerbo, Kristen},
  year={2024},
  publisher={Jul}
}

@inproceedings{gebreegziabher2025metricmate,
  title={MetricMate: An Interactive Tool for Generating Evaluation Criteria for LLM-as-a-Judge Workflow},
  author={Gebreegziabher, Simret Araya and Chiang, Charles and Wang, Zichu and Ashktorab, Zahra and Brachman, Michelle and Geyer, Werner and Li, Toby Jia-Jun and G{\'o}mez-Zar{\'a}, Diego},
  booktitle={Proceedings of the 4th Annual Symposium on Human-Computer Interaction for Work},
  pages={1--18},
  year={2025}
}

@inproceedings{desmond2025evalassist,
  title={Evalassist: Llm-as-a-judge simplified},
  author={Desmond, Michael and Ashktorab, Zahra and Geyer, Werner and Daly, Elizabeth M and Cooper, Martin Santillan and Pan, Qian and Nair, Rahul and Wagner, Nico and Pedapati, Tejaswini},
  booktitle={Proceedings of the AAAI Conference on Artificial Intelligence},
  volume={39},
  number={28},
  pages={29637--29639},
  year={2025}
}

@inproceedings{desmond2024evalullm,
  title={EvaluLLM: LLM assisted evaluation of generative outputs},
  author={Desmond, Michael and Ashktorab, Zahra and Pan, Qian and Dugan, Casey and Johnson, James M},
  booktitle={Companion proceedings of the 29th international conference on intelligent user interfaces},
  pages={30--32},
  year={2024}
}

@misc{Wiltberger_2025, title={Stop “Vibe testing” your LLMS. it’s time for real evals.}, url={https://developers.googleblog.com/en/streamline-llm-evaluation-with-stax/}, journal={Google Developers Blog}, publisher={Google Labs}, author={Wiltberger, Sara}, year={2025}, month={Aug}}

@article{ramnath2025amulet,
  title={Amulet: Putting Complex Multi-Turn Conversations on the Stand with LLM Juries},
  author={Ramnath, Sahana and Mudgil, Anurag and Joshi, Brihi and Hallinan, Skyler and Ren, Xiang},
  journal={arXiv preprint arXiv:2505.20451},
  year={2025}
}

@inproceedings{chiang2024chatbot,
  title={Chatbot arena: An open platform for evaluating llms by human preference},
  author={Chiang, Wei-Lin and Zheng, Lianmin and Sheng, Ying and Angelopoulos, Anastasios Nikolas and Li, Tianle and Li, Dacheng and Zhu, Banghua and Zhang, Hao and Jordan, Michael and Gonzalez, Joseph E and others},
  booktitle={Forty-first International Conference on Machine Learning},
  year={2024}
}

@inproceedings{kahng2024llm,
  title={Llm comparator: Visual analytics for side-by-side evaluation of large language models},
  author={Kahng, Minsuk and Tenney, Ian and Pushkarna, Mahima and Liu, Michael Xieyang and Wexler, James and Reif, Emily and Kallarackal, Krystal and Chang, Minsuk and Terry, Michael and Dixon, Lucas},
  booktitle={Extended Abstracts of the CHI Conference on Human Factors in Computing Systems},
  pages={1--7},
  year={2024}
}

@article{malouff2014preventing,
  title={Preventing halo bias in grading the work of university students},
  author={Malouff, John M and Stein, Sarah J and Bothma, Lodewicka N and Coulter, Kimberley and Emmerton, Ashley J},
  journal={Cogent Psychology},
  volume={1},
  number={1},
  pages={988937},
  year={2014},
  publisher={Taylor \& Francis}
}

@incollection{jarke2010requirements,
  title={Requirements engineering in complex domains},
  author={Jarke, Matthias and Klamma, Ralf and Pohl, Klaus and Sikora, Ernst},
  booktitle={Graph Transformations and Model-Driven Engineering: Essays Dedicated to Manfred Nagl on the Occasion of his 65th Birthday},
  pages={602--620},
  year={2010},
  publisher={Springer}
}

@book{norman2024things,
  title={Things that make us smart},
  author={Norman, Don},
  chapter={Chapter 3: The Power of Representation},
  year={2024},
  publisher={Diversion Books}
}

@article{hogarth1992order,
  title={Order effects in belief updating: The belief-adjustment model},
  author={Hogarth, Robin M and Einhorn, Hillel J},
  journal={Cognitive psychology},
  volume={24},
  number={1},
  pages={1--55},
  year={1992},
  publisher={Elsevier}
}

@inproceedings{koccak2018sequential,
  title={Sequential updating: A behavioral model of belief change},
  author={Ko{\c{c}}ak, Korhan},
  booktitle={Tech. Rep., Technical report, Working Paper},
  year={2018}
}

@article{kim2025evalet,
  title={Evalet: Evaluating Large Language Models by Fragmenting Outputs into Functions},
  author={Kim, Tae Soo and Lee, Heechan and Lee, Yoonjoo and Seering, Joseph and Kim, Juho},
  journal={arXiv preprint arXiv:2509.11206},
  year={2025}
}

@inproceedings{choi2025bloomintent,
  title={BloomIntent: Automating Search Evaluation with LLM-Generated Fine-Grained User Intents},
  author={Choi, Yoonseo and Kim, Eunhye and Kim, Hyunwoo and Park, Donghyun and Lee, Honggu and Kim, Jin Young and Kim, Juho},
  booktitle={Proceedings of the 38th Annual ACM Symposium on User Interface Software and Technology},
  pages={1--34},
  year={2025}
}

@inproceedings{zamfirescu2023johnny,
  title={Why Johnny can’t prompt: how non-AI experts try (and fail) to design LLM prompts},
  author={Zamfirescu-Pereira, J Diego and Wong, Richmond Y and Hartmann, Bjoern and Yang, Qian},
  booktitle={Proceedings of the 2023 CHI conference on human factors in computing systems},
  pages={1--21},
  year={2023}
}

@inproceedings{liu2022design,
  title={Design guidelines for prompt engineering text-to-image generative models},
  author={Liu, Vivian and Chilton, Lydia B},
  booktitle={Proceedings of the 2022 CHI conference on human factors in computing systems},
  pages={1--23},
  year={2022}
}

@article{peng2024survey,
  title={A survey of useful llm evaluation},
  author={Peng, Ji-Lun and Cheng, Sijia and Diau, Egil and Shih, Yung-Yu and Chen, Po-Heng and Lin, Yen-Ting and Chen, Yun-Nung},
  journal={arXiv preprint arXiv:2406.00936},
  year={2024}
}

@book{naik2011software,
  title={Software testing and quality assurance: theory and practice},
  author={Naik, Kshirasagar and Tripathy, Priyadarshi},
  year={2011},
  publisher={John Wiley \& Sons}
}

@inproceedings{yang2024generating,
  title={Generating Evaluation Criteria of Domain-Specific Large Language Model Using Word Vector Clustering},
  author={Yang, Jingyu and Xu, Hao and Wang, Rongxiao and Ming, Xuran and Li, Shoubin},
  booktitle={2024 IEEE 24th International Conference on Software Quality, Reliability, and Security Companion (QRS-C)},
  pages={94--100},
  year={2024},
  organization={IEEE}
}

@article{huovinen2024assessing,
  title={Assessing usability of large language models in education},
  author={Huovinen, Leo},
  year={2024}
}

@inproceedings{kim2023better,
  title={Which is better? exploring prompting strategy for llm-based metrics},
  author={Kim, Joonghoon and Lee, Sangmin and Han, Seung Hun and Park, Saeran and Lee, Jiyoon and Jeong, Kiyoon and Kang, Pilsung},
  booktitle={Proceedings of the 4th Workshop on Evaluation and Comparison of NLP Systems},
  pages={164--183},
  year={2023}
}

@article{ranjan2024comprehensive,
  title={A comprehensive survey of bias in llms: Current landscape and future directions},
  author={Ranjan, Rajesh and Gupta, Shailja and Singh, Surya Narayan},
  journal={arXiv preprint arXiv:2409.16430},
  year={2024}
}

@article{niu2025llama,
  title={Llama See, Llama Do: A Mechanistic Perspective on Contextual Entrainment and Distraction in LLMs},
  author={Niu, Jingcheng and Yuan, Xingdi and Wang, Tong and Saghir, Hamidreza and Abdi, Amir H},
  journal={arXiv preprint arXiv:2505.09338},
  year={2025}
}

@article{wu2025answer,
  title={Answer-Centric or Reasoning-Driven? Uncovering the Latent Memory Anchor in LLMs},
  author={Wu, Yang and Zhang, Yifan and Wang, Yiwei and Cai, Yujun and Wu, Yurong and Wang, Yuran and Xu, Ning and Cheng, Jian},
  journal={arXiv preprint arXiv:2506.17630},
  year={2025}
}

@inproceedings{liu2024calibrating,
  title={Calibrating llm-based evaluator},
  author={Liu, Yuxuan and Yang, Tianchi and Huang, Shaohan and Zhang, Zihan and Huang, Haizhen and Wei, Furu and Deng, Weiwei and Sun, Feng and Zhang, Qi},
  booktitle={Proceedings of the 2024 joint international conference on computational linguistics, language resources and evaluation (lrec-coling 2024)},
  pages={2638--2656},
  year={2024}
}

@inproceedings{amershi2019guidelines,
  title={Guidelines for human-AI interaction},
  author={Amershi, Saleema and Weld, Dan and Vorvoreanu, Mihaela and Fourney, Adam and Nushi, Besmira and Collisson, Penny and Suh, Jina and Iqbal, Shamsi and Bennett, Paul N and Inkpen, Kori and others},
  booktitle={Proceedings of the 2019 chi conference on human factors in computing systems},
  pages={1--13},
  year={2019}
}

@article{song2024human,
  title={Human-AI collaboration by design},
  author={Song, Binyang and Zhu, Qihao and Luo, Jianxi},
  journal={Proceedings of the Design Society},
  volume={4},
  pages={2247--2256},
  year={2024},
  publisher={Cambridge University Press}
}

@inproceedings{ashktorab2024trust,
  title={Trust and Reliance in Evolving Human-AI Workflows (TREW)},
  author={Ashktorab, Zahra and Bansal, Gagan and Bu{\c{c}}inca, Zana and Holstein, Kenneth and Hullman, Jessica and Smith-Renner, Alison Marie and Wu, Tongshuang and Zhang, Wenjuan},
  booktitle={Extended Abstracts of the CHI Conference on Human Factors in Computing Systems},
  pages={1--6},
  year={2024}
}

@inproceedings{wang2020human,
  title={From human-human collaboration to Human-AI collaboration: Designing AI systems that can work together with people},
  author={Wang, Dakuo and Churchill, Elizabeth and Maes, Pattie and Fan, Xiangmin and Shneiderman, Ben and Shi, Yuanchun and Wang, Qianying},
  booktitle={Extended abstracts of the 2020 CHI conference on human factors in computing systems},
  pages={1--6},
  year={2020}
}

@article{hymel2025analysis,
  title={Analysis of LLMs vs Human Experts in Requirements Engineering},
  author={Hymel, Cory and Johnson, Hiroe},
  journal={arXiv preprint arXiv:2501.19297},
  year={2025}
}

@inproceedings{ahmed2025can,
  title={Can llms replace manual annotation of software engineering artifacts?},
  author={Ahmed, Toufique and Devanbu, Premkumar and Treude, Christoph and Pradel, Michael},
  booktitle={2025 IEEE/ACM 22nd International Conference on Mining Software Repositories (MSR)},
  pages={526--538},
  year={2025},
  organization={IEEE}
}

@inproceedings{degachi2025towards,
  title={Towards a Domain Expert Evaluation Framework for Conversational Search in Healthcare},
  author={Degachi, Chadha and Dhar, Ujjayan and Niforatos, Evangelos and Kortuem, Gerd},
  booktitle={Proceedings of the Extended Abstracts of the CHI Conference on Human Factors in Computing Systems},
  pages={1--9},
  year={2025}
}

@inproceedings{sadek2024guidelines,
  title={Guidelines for integrating value sensitive design in responsible AI toolkits},
  author={Sadek, Malak and Constantinides, Marios and Quercia, Daniele and Mougenot, Celine},
  booktitle={Proceedings of the 2024 CHI Conference on Human Factors in Computing Systems},
  pages={1--20},
  year={2024}
}

@inproceedings{van2021towards,
  title={Towards guidelines for designing human-in-the-loop machine training interfaces},
  author={Van Der Stappen, Almar and Funk, Mathias},
  booktitle={Proceedings of the 26th International Conference on Intelligent User Interfaces},
  pages={514--519},
  year={2021}
}

\appendix
\section{Prompts}
\label{appendix:prompts}
This section presents the three scenarios used to develop evaluation criteria in both the nutrition domain as well as the math pedagogy domain. The same prompts were provided to the domain expert, LLM, and lay user, ensuring consistency in developing criteria across different perspectives.

\subsection{Nutrition and Dietetics Domain}
\label{appendix:nutritionprompts}

\raggedright
\textbf{Scenario 1.}
\label{appendix:nutscenario1}

\texttt{I'm aiming to cut down on added sugars and refined carbs to help manage my blood sugar levels, while also trying to increase my intake of fiber-rich foods.  I need something quick and easy in the morning, so I am considering starting the day with Great Value Cinnamon Crunch Breakfast Cereal. Please provide an explanation as to  why this product is a good or bad food choice for me, including any health benefits, allergens, dietary restrictions, or potential cereal substitutions.}\\[10pt]

\textbf{Scenario 2.}
\label{appendix:nutscenario2}

\texttt{I am a 25-year-old woman with Crohn's disease.  I have a lot of flare ups with the foods that I am eating.  Please discuss which foods I should be increasing in my diet and which ones I should be avoiding that could cause flare ups. If I follow these restrictions, please discuss what benefits I should be looking for.}\\[10pt]

\textbf{Scenario 3.}
\label{appendix:nutscenario3}

\texttt{I am a 35-year-old, 230 pound man that would like to lose weight and gain muscle. I have been going to the gym every day to workout. What are some diets that I should be considering to lose weight?}   
\raggedright

\subsection{Education and Tutoring Domain}
\label{appendix:tutoringprompts}

\raggedright
\textbf{Scenario 1.}
\label{appendix:edscenario1}

\texttt{I am a freshman in high school studying Algebra 1, and I'm working through a homework problem that's giving me some trouble. Here's the problem I'm working on:} \\

\texttt{Solve for \( x \) in the equation:}\\
\[
\frac{3x - 2}{4} + 5 = \frac{7}{2}.
\]
\\
\texttt{So far, I have subtracted 5 from both sides of the equation to get:}\\
\[
\frac{3x - 2}{4} = \frac{7}{2} - 5.
\]
\\
\texttt{I'm confused about how to deal with the fractions and what to do next. Could you please give me some hints on how to continue without directly giving me the answer? }
\\[10pt]

\textbf{Scenario 2.}
\label{appendix:edscenario2}

\texttt{I am a 9th-grade student and currently learning about relative speed and rate problems in my Algebra class. Could you please review my solution and provide feedback? I'd appreciate it if you could point out any errors in my reasoning or suggest alternative approaches to solving this problem. I'm really trying to grasp this concept fully.} \\[10pt]
\texttt{Question:} 
\texttt{Car A is traveling from City X to City Y, a distance of 120 miles, at a constant speed of 40 miles per hour. Car B is traveling from City Y to City X, a distance of 120 miles, at a constant speed of 60 miles per hour. Both cars start at the same time. At what point along the road between City X and City Y will they meet?} \\[10pt]
\texttt{My Solution:} 
\texttt{First, I set the distances they travel equal to each other since they meet at the same point: 40t = 60t Then, I solve for t by dividing both sides by their respective speeds, so t = 40/60 Therefore, they will meet at t = 2/3 hours.} \\[10pt]

\textbf{Scenario 3}
\label{appendix:edscenario3}

\texttt{I am a high school student in a math class. Can you help me understand how to solve quadratic equations using the quadratic formula? I'm not sure when or why to use it.}
\raggedright

\section{LLM Criteria Generation Prompts}
\label{appendix:criteria_generation}

This section presents the single prompt used to develop the evaluation criteria. It includes one prompt for generating criteria based solely on the initial instruction, and another for refining the criteria after considering the model's outputs. Blue text indicates content that is programmatically filled in.

\subsection{Criteria Generation Based on Instruction Only}

\textbf{System Prompt}             
You are a helpful and precise assistant that can create binary evaluation criteria for a given user instruction. Your task is to generate evaluation criteria for assessing a large language model's performance. Each criterion should be a statement in which you would answer true/false. The criteria should be in the form of a one sentence statement, not a question. You should return your final answer as a valid JSON object.

\textbf{User Prompt} Create evaluation criteria for the given prompt instruction. \\
\vspace{0.5\baselineskip} 

\textit{[The Start of Instructions]} \\
\textcolor{blue}{Instruction} \\
\textit{[The End of Instructions]} \\[10pt]

\subsection{Criteria Refinement Incorporating Model Outputs}
\textbf{System Prompt} 
You are a helpful and precise assistant that will refine or add new binary evaluation criteria for a given user instruction and set out outputs. Your task is to refine or create new evaluation criteria for assessing a large language model's performance given a set of initial evaluation criteria. Each criterion should be a statement in which you would answer true/false. The criteria should be in the form of a one sentence statement, not a question. The response should separate out the revised criteria and added criteria from the initial criteria.  You should return your final answer as a valid JSON object.

\textbf{User Prompt} Consider the initial evaluation criteria provided. After reviewing the outputs attached, either refine the intial criteria or create new evaluation criteria for the given prompt instruction and three outputs: \\
\textit{[The Start of Criteria]} \\
\textcolor{blue}{Evaluation Criteria} \\
\textit{[The End of Criteria]} \\[10pt]

After reviewing the outputs attached, either refine the initial criteria or create new evaluation criteria for the given prompt instruction and three outputs. \\[10pt]

\textit{[The Start of Instructions]} \\
\textcolor{blue}{Instruction} \\
\textit{[The End of Instructions]} \\[10pt]

\textit{[The Start of Assistant 1’s Response]} \\
\textcolor{blue}{Output 1} \\
\textit{[The End of Assistant 1’s Response]} \\[10pt]

\textit{[The Start of Assistant 2’s Response]} \\
\textcolor{blue}{Output 2} \\
\textit{[The End of Assistant 2’s Response]} \\[10pt]

\textit{[The Start of Assistant 3’s Response]} \\
\textcolor{blue}{Output 3} \\
\textit{[The End of Assistant 3’s Response]} \\[10pt]


\end{document}